\documentclass{aa}

\usepackage{graphicx}
\usepackage{txfonts}

\usepackage{caption}
\usepackage{subcaption}

\usepackage{natbib}
\usepackage{amsmath}
\usepackage{hyperref}

\newcommand{\coord}[2]{(#1,\,#2)}

\begin{document}

\title{From star-disc encounters to numerical solutions for a subset of the restricted three-body problem}

\author{Andreas Breslau
  \and Kirsten Vincke
  \and Susanne Pfalzner
}

\institute{Max-Planck-Institut f\"ur Radioastronomie, Auf dem H\"ugel 69, 53121 Bonn, Germany\\
  \email{abreslau@mpifr.de}
}

\date{ }

\abstract{Various astrophysical processes exist, where the fly-by of a massive object affects matter that is initially supported against gravity by rotation.
  Examples are perturbations of galaxies, protoplanetary discs, or planetary systems.
  We approximate such events as a subset of the restricted three-body problem by considering only
  perturbations of non-interacting low-mass objects that are initially on circular Keplerian orbits.
  In this paper, we present a new parametrisation of the initial conditions of this problem.
  Under certain conditions, the initial positions of the low-mass objects
  can be specified as being largely independent of the initial position of the perturber.
  In addition, exploiting the known scalings of the problem
  reduces the parameter space of initial conditions for one specific perturbation to two dimensions.
  To this two-dimensional initial condition space, we have related the final properties of the perturbed trajectories
  of the low-mass objects from our numerical simulations.
  In this way, maps showing the effect of the perturbation on the low-mass objects were created,
  which provide a new view on the perturbation process.
  Comparing the maps for different mass-ratios reveals that the perturbations by low- and high-mass perturbers are
  dominated by different physical processes. The equal-mass case is a complicated mixture of the other two cases.
  Since the final properties of trajectories with similar initial conditions are also usually similar,
  the results of the limited number of integrated trajectories can be generalised to the full presented parameter space
  by interpolation.
  Since our results are also unique within the accuracy strived for, they constitute general numerical solutions for this subset of the restricted three-body problem.
  As such, they can be used to predict the evolution of real physical problems by simple transformations, such as scaling,
  without further simulations.
  Possible applications are the perturbation of protoplanetary discs or planetary systems by the fly-by of another star.
  Here, the maps enable us, for example, to quantify the portion of unbound material for any periastron distance
  without the need for further simulations.
}


\maketitle

{\bf
After publication of this paper\footnote{\bf DOI: 10.1051/0004-6361/201526068},
we noticed a mistake in the plot script used to produce Figures~2d, 3d, and 5d.
To correct this mistake, we published a Corrigendum\footnote{\bf DOI: 10.1051/0004-6361/201526068e}.
In this update of the preprint, we replaced the wrong images and text with the corrected ones (corrected text highlighted in bold).
}
 
\section{Introduction}
\label{sec:intro}

Various astrophysical processes exist where matter, initially supported against gravity by rotation,
is affected by the gravitational force of a passing-by massive object.
Among them are, for example, galaxy-galaxy encounters and perturbations of protoplanetary discs, debris discs, or planetary systems by passing stars
\citep[see e.g.][]{1972ApJ...178..623T,1981ApJ...243...32F,1993MNRAS.261..190C,1995ApJ...455..252H, 1996MNRAS.278..303H, 1997MNRAS.287..148H, 2001Icar..153..416K,2002ESASP.500..305M,2004AJ....128.2553L, 2005A&A...437..967P, 2006ApJ...642.1140O,2007A&A...469..913S,2011MNRAS.411..859M, 2013ApJ...769..150C,2014MNRAS.444.2808P}.
Some of these processes happen on different scales and might involve additional forces, such as viscous friction or magnetic fields.
Nevertheless, what they all have in common is that, approximately at the closest approach of the perturber,
the perturber's gravitational force dominates the event.
Therefore, it is very likely that, at least, matter that is close to the perturber at pericentre passage
is in all cases affected in a similar way.

For simplicity, we focus here on cases where relatively low-mass matter is orbiting a massive object (the host)
and is perturbed by another massive object (the perturber).
As examples, these can be a single planet, or a disc of gas or debris orbiting a star
that is perturbed by another star flying by.
In the case of a disc, we follow the approach of several previous studies, \citep[e.g.][]{1972ApJ...178..623T,1996MNRAS.278..303H, 2001Icar..153..416K,2005A&A...437..967P}
and neglect interactions of the disc material with itself, such as self-gravity or viscosity, to approximate all these cases as a subset of the restricted three-body problem.

The investigation of the three-body problem has a long history starting with
Newton, Euler, Lagrange, and Jacobi (see e.g. \citealp{1990tbp..book.....M,2006tbp..book.....V,2014RPPh...77f5901M} and references therein).
A well studied subset of the general three-body problem is the perturbation of stable two-body systems by a third body.
This problem was investigated analytically \citep[e.g.][]{1983ApJ...268..342H, 1993ApJS...85..347H}
and numerically \citep[e.g.][]{1979CeMec..19...53V, 1983ApJ...268..319H, 1993ApJS...85..347H}.
The perturbation of low-mass matter owing to a fly-by is related to this problem.
It was analytically approximated, for example, by \citet{1994ApJ...424..292O}, \citet{1997MNRAS.290..490L}, and \citet{2010ApJ...725..353D}.
However, to date, no satisfying general analytical description of such perturbations has been found.
Therefore, previous studies investigating the influence of an encounter on planetary systems or the
global properties of discs (e.g. energy, mass, size)
usually performed statistical numerical simulations
that cover certain subsets of the large space of encounter parameters
\citep[e.g.][]{1996MNRAS.278..303H, 1997MNRAS.287..148H, 2005A&A...437..967P, 2006ApJ...640.1086F, 2012A&A...538A..10S, 2013MNRAS.433..867H, 2014A&A...565A.130B, 2015MNRAS.448..344L}.

Also, based on numerical simulations, some approaches were made to generalise the effect of these perturbations.
As such, usually the properties of the perturbed system were related to initial properties.

\citet{1993MNRAS.261..190C} and \citet{1996MNRAS.278..303H} investigated the perturbation of protoplanetary discs in equal-mass encounters 
and presented the probability for a disc particle to have a certain fate (remaining bound, captured by perturber, unbound) depending on its initial distance
to the host, $r_{\mathrm{init}} $, normalised to the perturber's periastron distance, $r_{\mathrm{p,peri}} $.
\footnote{
  Variables with the index 'p' denote properties of the perturber or its orbit.
  Variables without this index usually refer to the particle.
}
The perturber's periastron distance here and in the following means the distance between host and perturber
at the moment when the perturber is in the pericentre of its orbit.

In a similar scenario, \citet{1996MNRAS.278..303H} identified the particles that undergo the largest energy change during an encounter
and located them within the disc (see their Figs.~4a, 5, \mbox{and 6).}
They found that, shortly before periastron passage,
these particles formed four distinct groups that interact with the host and perturber in different ways.
However, since the disc is already deformed at the time of periastron passage, a description of the spatial distribution of these particles proved difficult.

\citet{1972ApJ...178..623T} performed simulations of galaxy perturbations with non-interacting low-mass disc particles.
Since they modelled the gravitational potential of the involved galaxies by point masses,
their results are comparable to simulations of low-mass discs around stars.
Similar to \citet{1993MNRAS.261..190C} and \citet{1996MNRAS.278..303H}, they showed the fates (remaining bound,
captured by perturber, unbound, forming tails, and a bridge between host and perturber) of the disc particles,
but, depending on ``the positions they would have reached at time t = 0 in the absence of forcing''.
By ``absence of forcing'', they mean without the force from the perturber on the particles and in their setup,
$t = 0$ is the moment of the perturbers pericentre passage (see their Fig.~15 and the describing text).

Using a similar approach, like \citet{1972ApJ...178..623T},
we developed a new parametrisation of the initial conditions of the problem.
For a given perturbation
\footnote{A certain perturbation is here defined by the mass of the perturber relative to the mass of the host,
  $m$, the eccentricity, $e_{\mathrm{p}}$, of the perturber orbit,
  and the orientation of the plane of the perturber orbit relative to the plane of the orbiting low-mass object.
}, the resulting two-dimensional initial condition space
enables the creation of maps of the properties of the system, e.g. the final values.
This new view on the problem provides a better understanding of the underlying perturbation process itself.
Since the solution for one set of initial conditions is unique,
once obtained, numerical results can be easily applied to different physical problems
by simple transformations, such as scaling, and thus making additional simulations unnecessary.
This reduces the computational effort of parameter studies in this context significantly.

This paper focuses on the presentation of the new method,
which is described in Sect.~\ref{sec:method}, along with the numerical scheme that was used.
In Sect.~\ref{sec:results}, some exemplary results of the new method are provided in the form of the above-mentioned maps.
The limitations of the method along with the application to real physical systems are discussed in Sect.~\ref{sec:discussion}.

\section{Method}
\label{sec:method}

Our new method is basically a refinement of the approach used by \citet{1972ApJ...178..623T} for their Fig.~15.
Since they mentioned it only in passing, and without complete description,
we first explain their approach and then describe how we developed it further.

To describe it simply, we consider a disc of mass-less particles
initially orbiting counter-clockwise on circular Keplerian orbits in the $x$-$y$-plane
around a host with mass $M_{\mathrm{h}} $ in the centre of the coordinate system.
These particles are perturbed by a perturber of mass $M_{\mathrm{p}} $.
The actual orbit of the perturber depends on the mass-ratio of host and perturber, \mbox{$m = M_{\mathrm{p}} /M_{\mathrm{h}} $},
the eccentricity of the orbit, $e_{\mathrm{p}}$,
the pericentre distance, $r_{\mathrm{p,peri}} $, and the plane of the orbit relative to the orbit of the particles.
We further restrict this to cases where the orbit of the particles and the orbit of the perturber
are coplanar and prograde, thus reducing the problem to two dimensions.
The line from the host to the pericentre of the perturber's orbit defines the $x$-axis of our reference system.
The particles' gravitational effect on host and perturber is assumed to be negligible.

As in some previous studies \citep[e.g.][]{1972ApJ...178..623T,1993MNRAS.261..190C,1996MNRAS.278..303H},
we scale all lengths of the problem by setting $r_{\mathrm{p,peri}}  = 1$.
The initial radius of particle $i$'s orbit, $r_{i \mathrm{,init}} $, is then given in units of $r_{\mathrm{p,peri}} $.

We assume the perturber to pass the host on a parabolic orbit. Therefore, their distance is
at $t = -\infty$ and $t = \infty$ in principle infinite.
In the moment of pericentre passage the distance is smallest, namely $r_{\mathrm{p,peri}} $, and
the perturber's influence on the host is strongest.
The gravitational force of the perturber on the particle is strongest close to pericentre passage.
Since it is impossible to cover the period from $t=-\infty$ to $t=\infty$ in a simulation, this
fact can be used to restrict the investigation of the perturbation of the particle's orbit
to a time period around pericentre passage
\footnote{
  Indeed, this fact is used in all numerical investigations of gravitational perturbations
  to restrict the simulation time
  (e.g. \citet{1996MNRAS.278..303H} start their investigation of the perturbation of a disc
  with an initial distance of at least $10$ times the disc radius).
}.
The perturbed particle trajectories obtained for one specific perturber orbit
converge with increasing initial separation between perturber and host.

Owing to this convergence, numerical investigations of the perturbation process should start
when the perturber's influence on the particles is still negligible.
At that time ($t = t_{\mathrm{init}} $), the perturber's position is given in polar coordinates by \mbox{($r_{\mathrm{p,init}}$, $\theta_{\mathrm{p,init}}$)}.
Owing to the above's condition, the perturber is then far outside the initial particle orbits, $r_{\mathrm{p,init}} \gg r_{i \mathrm{,init}} $.
For a parabolic orbit the time of flight from this initial position to pericentre is given by
\footnote{
  The mathematical bases can be found for example in \cite{1971fuas.book.....B} or Appendix 1.5 of \cite{2006imad.book.....K}.
}
\begin{align}
  t_{\mathrm{peri}}  - t_{\mathrm{init}}  = \sqrt{\frac{ 2 r_{\mathrm{p,peri}} ^3 }{ \mu } } \left(E - e_{\mathrm{p}} \sin{E} \right)\label{eq:time_of_flight}
\end{align}
with the standard gravitational parameter for the two-body problem $\mu = G (M_{\mathrm{h}}  + M_{\mathrm{p}} )$ and the eccentric anomaly
\begin{align}
  E = \arccos{\frac{e_{\mathrm{p}} - \cos{\theta_{\mathrm{p,init}} } }{ 1 + e_{\mathrm{p}} \cos{\theta_{\mathrm{p,init}}} } }.\label{eq:eccentric_anomaly}
\end{align}
Since the particles orbit the host initially on circular Keplerian orbits with radii $r_{i \mathrm{,init}} $,
their initial angular velocities are given by
\begin{align}
  \omega_{i \mathrm{,0}}  = \sqrt{G \frac{M_{\mathrm{h}} }{r_{i \mathrm{,init}} ^3}},\label{eq:omega}
\end{align}
where $G$ is the gravitational constant.
The particles' orbital periods are $T_{i} = 2\pi /\omega_{i \mathrm{,0}} $.
Without the influence of the perturber, the positions of the particles are given for all times
by the initial radius $r_{i \mathrm{,init}} $ and the angle
\begin{align}
  \theta_{i} (t) = \theta_{i \mathrm{,0}}  + t \, \omega_{i \mathrm{,0}} .\label{eq:thetai}
\end{align}
During the period which the perturber needs to reach pericentre ($t_{\mathrm{peri}}  - t_{\mathrm{init}} $),
a particle that is initially (at \mbox{$t = t_{\mathrm{init}} $}) at the position ($\theta_{i \mathrm{,init}} $,$r_{i \mathrm{,init}} $),
would move without the perturber's influence to ($\theta_{i \mathrm{,init}}  + \omega_{i \mathrm{,0}}  \, (t_{\mathrm{peri}}  - t_{\mathrm{init}} )$, $r_{i \mathrm{,init}} $)
\footnote{
  This position should not be confused with \mbox{($\theta_{i} (t_{\mathrm{peri}} )$, $r_{i}(t_{\mathrm{peri}} )$)},
  where the particle will be at $t = t_{\mathrm{peri}} $ with the perturbation.
}.
This is the position \citet{1972ApJ...178..623T} called ``the position [it] would have reached at time t = 0 in the absence of forcing”.
Since the particle will never be at this position, we will call it virtual pericentre position (VPP) in the following.
This position is obtained from the initial position by the transformation
\begin{align}
  \theta_{i \mathrm{,vpp}}  &= \left[ \theta_{i \mathrm{,init}}  + \omega_{i \mathrm{,0}}  \, (t_{\mathrm{peri}}  - t_{\mathrm{init}} ) \right]\mod 2 \pi\label{eq:tvpp}\\
  r_{i \mathrm{,vpp}}  &= r_{i \mathrm{,init}} .\label{eq:rvpp}
\end{align}

\begin{figure}[t!]
  \centering
  \begin{minipage}[t]{\hsize}
    \vspace{0pt}
    \begin{subfigure}[t]{\textwidth}
      \begin{minipage}[t]{0.05\textwidth}
        \vspace{0pt}
        \caption{}\label{fig:unspiral_1}
      \end{minipage}
      \hfill
      \begin{minipage}[t]{0.94\textwidth}
        \vspace{0pt}
        \centering
        \includegraphics[width=0.8\textwidth]{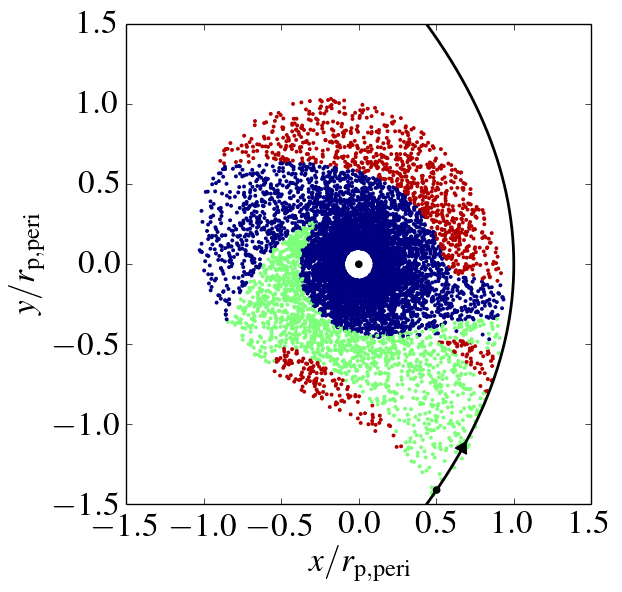}
      \end{minipage}
    \end{subfigure}
  \end{minipage}
  \hspace{1em}
  \begin{minipage}[t]{\hsize}
    \vspace{0pt}
    \begin{subfigure}[t]{\textwidth}
      \begin{minipage}[t]{0.05\textwidth}
        \vspace{0pt}
        \caption{}\label{fig:unspiral_2}
      \end{minipage}
      \hfill
      \begin{minipage}[t]{0.94\textwidth}
        \vspace{0pt}
        \centering
        \includegraphics[width=0.8\textwidth]{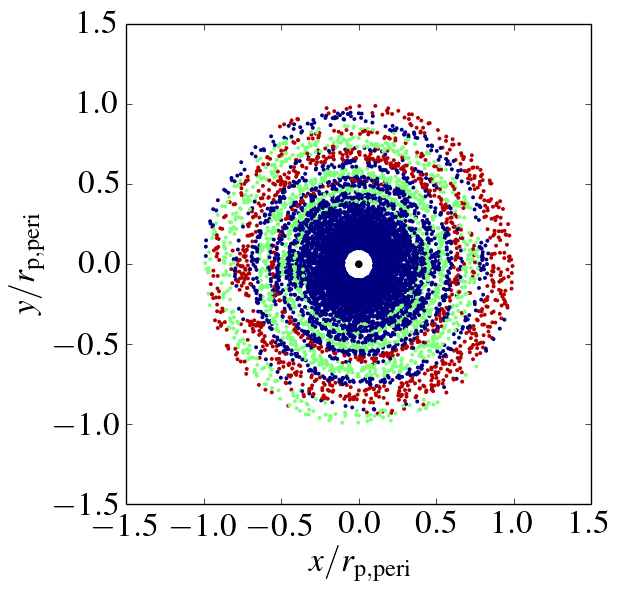}
      \end{minipage}
    \end{subfigure}
  \end{minipage}
  \hspace{1em}
  \begin{minipage}[t]{\hsize}
    \vspace{0pt}
    \begin{subfigure}[t]{\textwidth}
      \begin{minipage}[t]{0.05\textwidth}
        \vspace{0pt}
        \caption{}\label{fig:unspiral_3}
      \end{minipage}
      \hfill
      \begin{minipage}[t]{0.94\textwidth}
        \vspace{0pt}
        \centering
        \includegraphics[width=0.8\textwidth]{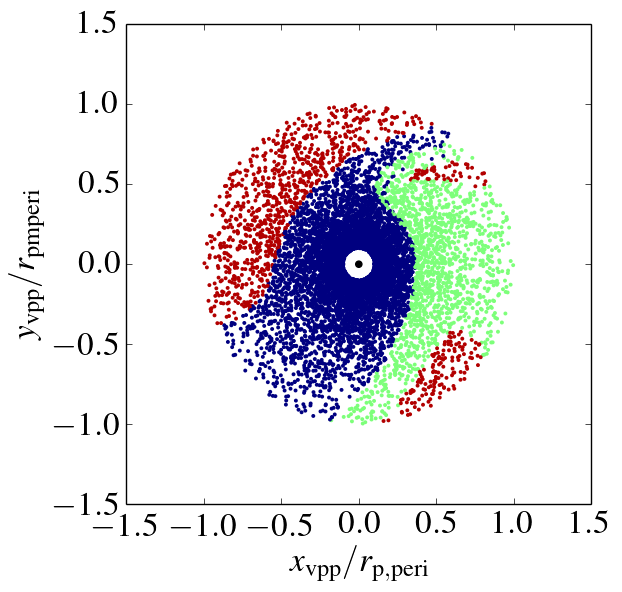}
      \end{minipage}
    \end{subfigure}
  \end{minipage}
  \caption{Virtual pericentre positions (VPPs) for a perturbed disc.
    The perturber has the same mass as the host and the pericentre of the parabolic, prograde orbit is at (1,0).
    In {\bf a)} and {\bf b)} the disc is seen face on when the perturber has nearly reached pericentre and at $t = t_{\mathrm{init}} $, respectively.
    In {\bf c)} the VPPs of the disc particles are shown. For full description see text.}
  \label{fig:unspiral}
\end{figure}
Figure~\ref{fig:unspiral} shows an example of this transformation for a disc of particles perturbed by a perturber with mass $M_{\mathrm{p}}  = M_{\mathrm{h}} $.
The perturber passes the host on a parabolic orbit coming from the bottom of the image,
passing pericentre at (1,0), and leaving at the top of the image (see black line in Fig.~\ref{fig:unspiral_1}).
The particles are coloured by their final fates: blue particles remain bound to the host,
green particles become captured by the perturber, and red particles become unbound.
Figure~\ref{fig:unspiral_1} shows the particles when the perturber has nearly reached
pericentre (perturber position indicated by a black dot at the bottom of the image).
Owing to the influence of the perturber, the disc is already deformed at that moment.

Figure~\ref{fig:unspiral_2} shows the same for $t= t_{\mathrm{init}} $.
Because of the differential rotation within the disc, the pattern from Fig.~\ref{fig:unspiral_1} appears ``wound up``.
In contrast to Fig.~\ref{fig:unspiral_1}, the disc is not yet deformed by the influence of the perturber.

Figure~\ref{fig:unspiral_3} shows the particles at their VPPs.
In this case, the disc is neither deformed nor is the pattern from Fig.~\ref{fig:unspiral_1} wound up
like in Fig.~\ref{fig:unspiral_2}.

\subsection{New parametrisation of the initial conditions}
\label{sec:vpp}

Instead of setting up simulations by sampling the initial positions of the particles (at $t = t_{\mathrm{init}} $) randomly
and plotting the particles later at the VPPs, as shown for example in Fig.~\ref{fig:unspiral_3},
we invert the transformation given by Eqs.~\eqref{eq:tvpp} and \eqref{eq:rvpp} to parametrise the initial conditions.
This is analogous to the definition of the impact parameter for the description of scattering processes,
e.g. of a test-particle in the field of a single attracting body.
The impact parameter describes at which distance the test-particle would pass the target without being influenced.
Similarly, we define the initial particle positions by the positions where the particles would be when the perturber is in pericentre
without the perturbation.
For a given initial distance between host and perturber, the initial particle positions are obtained by sampling the VPP space 
and transforming the coordinates with
\begin{align}
  \theta_{i \mathrm{,init}}  &= \left[ \theta_{i \mathrm{,vpp}}  - \omega_{i \mathrm{,0}}  \, (t_{\mathrm{peri}}  - t_{\mathrm{init}} ) \right]\mod 2 \pi\label{eq:tinit}\\
  r_{i \mathrm{,init}}  &= r_{i \mathrm{,vpp}} .\label{eq:rinit}
\end{align}

\subsection{The numerical simulations}
\label{sec:numerics}

For our simulations, we sampled the VPP space spanned by $x_{\mathrm{vpp}} $ and $y_{\mathrm{vpp}} $ with massless tracer particles.
Both dimensions were divided into bins and sampled with particles, each in the centre of the respective bin.
For the figures presented in Sect.~\ref{sec:results}, the region $x_{\mathrm{vpp}} /r_{\mathrm{p,peri}}  \in [-5;5]$, $y_{\mathrm{vpp}} /r_{\mathrm{p,peri}}  \in [-5;5]$
was sampled with \mbox{$500 \cdot 500$} particles.
The mass of the host is $M_{\mathrm{h}}  = 1$, the perturber with mass $M_{\mathrm{p}}  = m$ moves on a parabolic orbit
($e_{\mathrm{p}} = 1$) with a pericentre of \mbox{$r_{\mathrm{p,peri}}  = 1$}.
The trajectory of each particle was integrated individually with the adaptive LSODA integrator from the ODEPACK library \citep{ODEPACK}.

The initial position of the perturber was determined individually for each particle.
To fulfil the claim that the initial influence of the perturber is negligible, the initial distance of the perturber was chosen so
that the force from the perturber onto the particle, $F_{\mathrm{p}}$, was small relative to the force from the host, $F_{\mathrm{h}}$:
\begin{align}
  F_{\mathrm{p}}/F_{\mathrm{h}} < \epsilon\label{eq:force}.
\end{align}

This maximum initial influence of the perturber determines the initial position of the perturber.
For a particle with \mbox{$r_{\mathrm{vpp}}  = \sqrt{x_{\mathrm{vpp}} ^2 + y_{\mathrm{vpp}} ^2}$}, the minimum initial distance between particle and perturber, $d_{\mathrm{p}}$, follows from Eq.~\eqref{eq:force}:
\begin{align}
  d_{\mathrm{p}} = r_{\mathrm{vpp}}  \sqrt{m / \epsilon}.
\end{align}
For practical reasons, the initial distance between host and perturber is set to $r_{\mathrm{p,init}} = d_{\mathrm{p}} + r_{\mathrm{vpp}} $.
From $r_{\mathrm{p,init}}$ and the eccentricity of the perturber orbit, $e_{\mathrm{p}}$, follows $\theta_{\mathrm{p,init}}$ according to
\begin{align}
  r_{\mathrm{p,init}} = \frac{r_{\mathrm{p,peri}}  (1 + e_{\mathrm{p}})}{1+ e_{\mathrm{p}} \cos(\theta_{\mathrm{p,init}})}
\end{align}
and thus the time until pericentre passage, $t_{\mathrm{peri}}  - t_{\mathrm{init}} $, with Eqs.~\eqref{eq:time_of_flight} and \eqref{eq:eccentric_anomaly}.
After determining the time until pericentre passage of the perturber, the initial position of the particle was obtained according to Eqs.~\eqref{eq:tinit} and \eqref{eq:rinit}.

Tests have shown that for $\epsilon = 10^{-4}$ the particles are sufficiently unperturbed at the onset of the simulations.
The obtained accuracy is discussed in Sect.~\ref{sec:discussion}.
With these initial conditions, the perturber is initially at much larger distances from the host than in most simulations of this kind that have been performed so far
since in earlier studies the computational resources necessary for a larger number of these type of simulations were not available.

During the simulation, it was regularly tested whether the particle had settled into a stable orbit.
Here, possible stable, final orbits are either bound orbits ($e \leq 1$) around host or perturber, or unbound orbits ($e > 1$) with the centre of mass.
A bound orbit with the centre of mass is only a transition state.
An orbit was accepted as stable when one of the above conditions was fulfilled and 
the orbital elements (eccentricity, semi-major axis, argument of periapsis, etc.) that were calculated for the respective orbit did not change more than $1\%$
between two tests.
The duration between two tests was usually $(t_{\mathrm{peri}} -t_{\mathrm{init}} ) / 50$.
When the orbit was stable, the integration of the particle's trajectory was stopped.

\section{Results}
\label{sec:results}

\begin{figure*}[h!t]
  \centering
  \begin{minipage}[t]{\hsize}
    \centering
    \begin{minipage}[t]{0.42\hsize}
      \vspace{0pt}
      \begin{subfigure}[t]{\textwidth}
        \begin{minipage}[t]{0.05\textwidth}
          \vspace{0pt}
          \caption{}\label{fig:m01_5x5_fates}
        \end{minipage}
        \hfill
        \begin{minipage}[t]{0.94\textwidth}
          \vspace{0pt}
          \includegraphics[width=\textwidth]{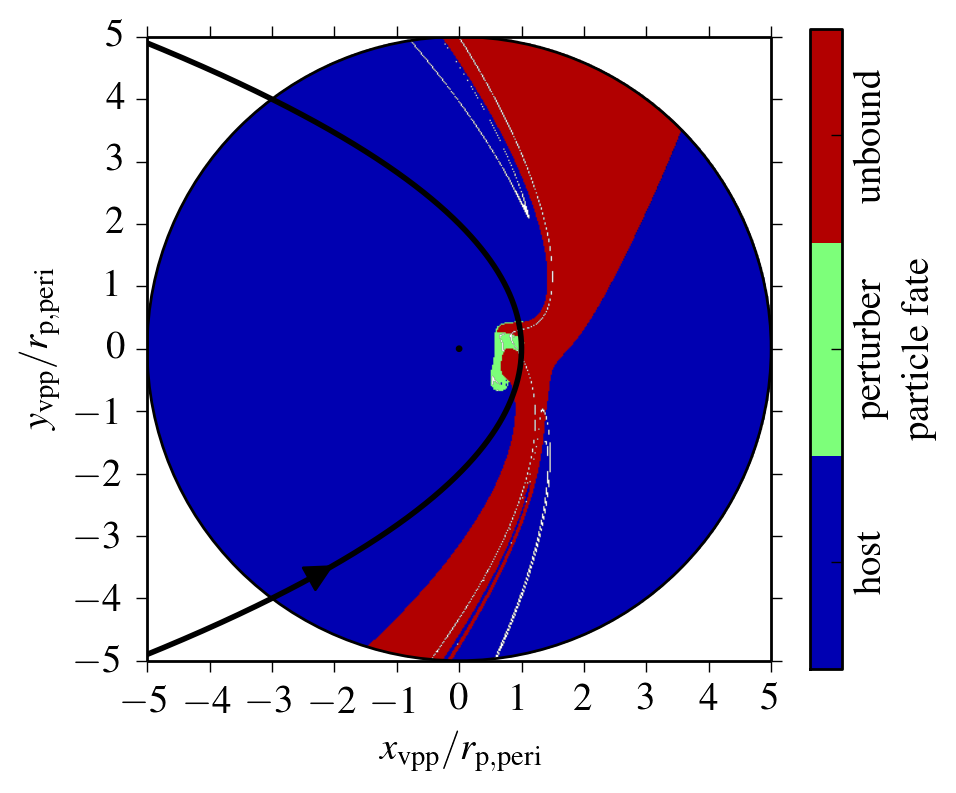}
        \end{minipage}
      \end{subfigure}
    \end{minipage}
    \hspace{1em}
    \begin{minipage}[t]{0.42\hsize}
      \vspace{0pt}
      \begin{subfigure}[t]{\textwidth}
        \begin{minipage}[t]{0.05\textwidth}
          \vspace{0pt}
          \caption{}\label{fig:m01_5x5_fates_io}
        \end{minipage}
        \hfill
        \begin{minipage}[t]{0.94\textwidth}
          \vspace{0pt}
          \includegraphics[width=\textwidth]{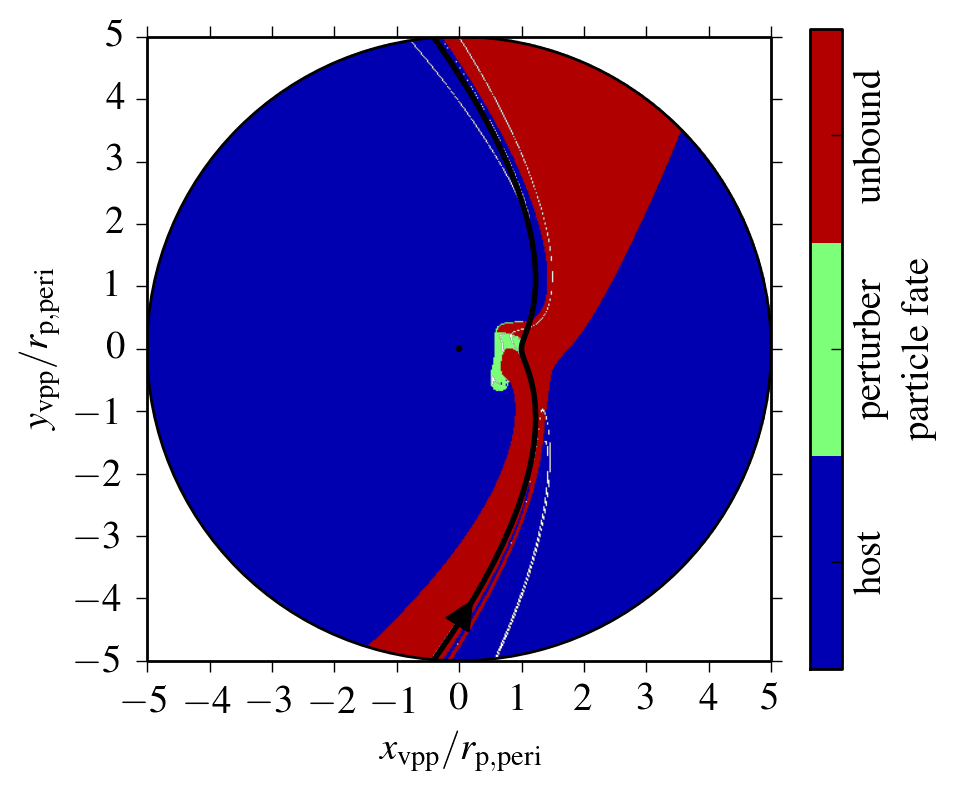}
        \end{minipage}
      \end{subfigure}
    \end{minipage}
  \end{minipage}
  \begin{minipage}[t]{\hsize}
    \centering
    \begin{minipage}[t]{0.42\hsize}
      \vspace{0pt}
      \begin{subfigure}[t]{\textwidth}
        \begin{minipage}[t]{0.05\textwidth}
          \vspace{0pt}
          \caption{}\label{fig:m01_5x5_ecc_bound}
        \end{minipage}
        \hfill
        \begin{minipage}[t]{0.94\textwidth}
          \vspace{0pt}
          \includegraphics[width=\textwidth]{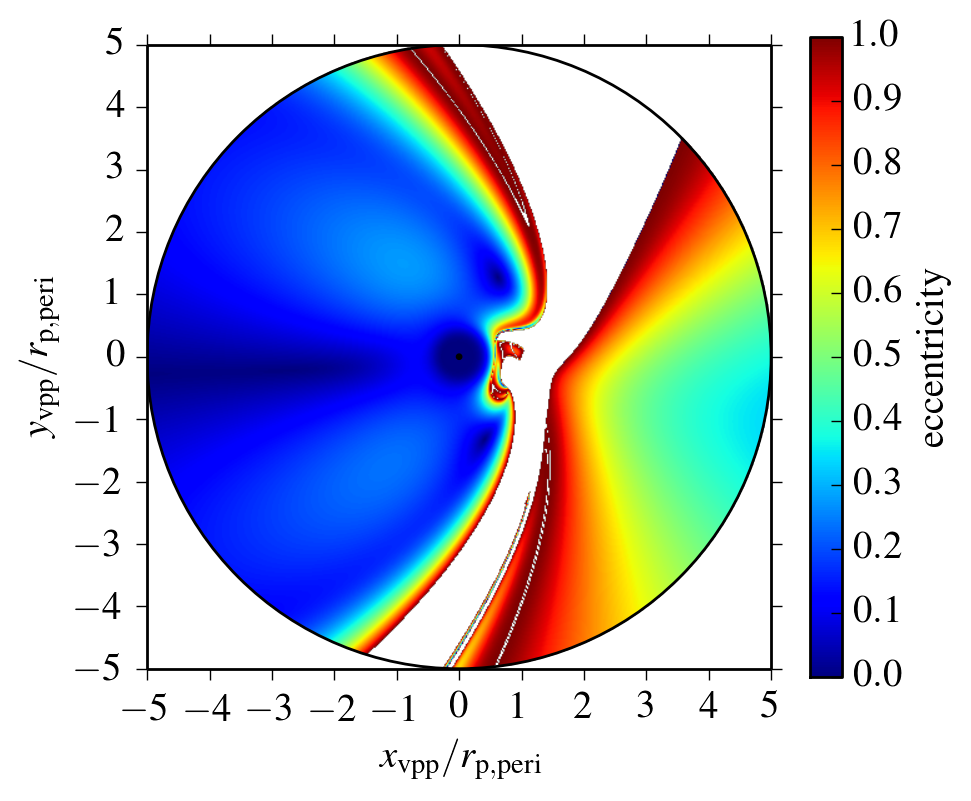}
        \end{minipage}
      \end{subfigure}
    \end{minipage}
    \hspace{1em}
    \begin{minipage}[t]{0.42\hsize}
      \vspace{0pt}
      \begin{subfigure}[t]{\textwidth}
        \begin{minipage}[t]{0.05\textwidth}
          \vspace{0pt}
          \caption{}\label{fig:m01_5x5_semimajor_bound}
        \end{minipage}
        \hfill
        \begin{minipage}[t]{0.94\textwidth}
          \vspace{0pt}
          \includegraphics[width=\textwidth]{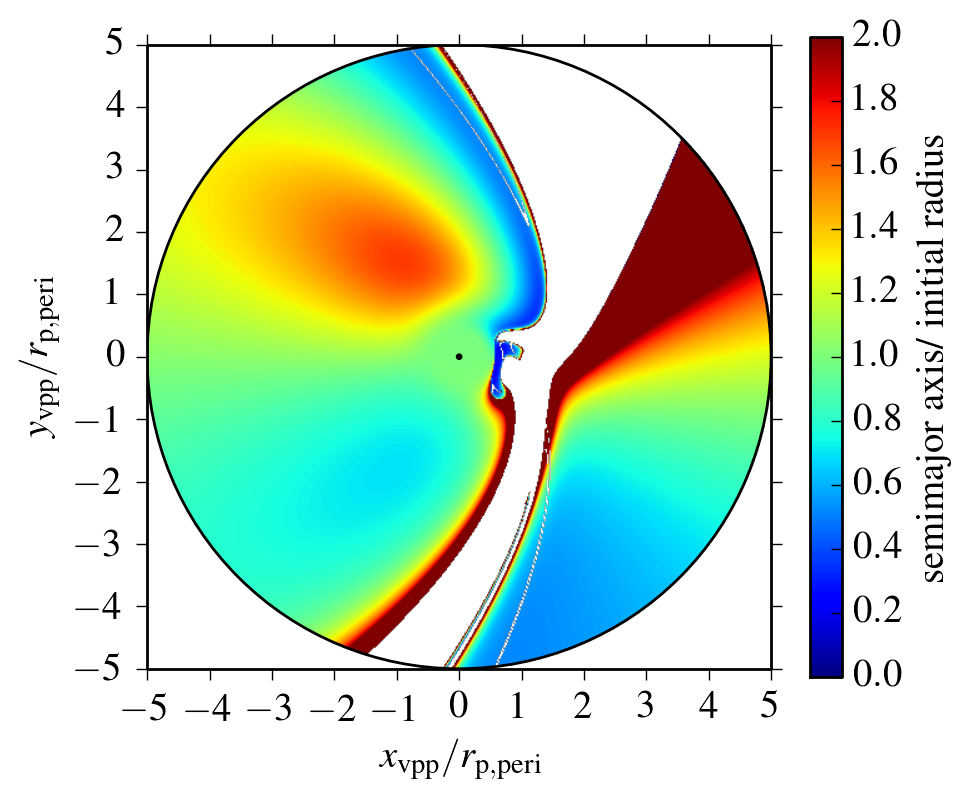}
        \end{minipage}
      \end{subfigure}
    \end{minipage}
  \end{minipage}
  \caption{Maps of the final properties of the particles depending on their VPPs for the region
    $r_{\mathrm{vpp}} /r_{\mathrm{p,peri}}  < 5 $ for a mass-ration of $m = 0.1$.\\
    {\bf a)} and {\bf b)} show whether the particles are finally bound to host or perturber, or become unbound.
    {\bf a)} also shows the orbit of the perturber through the Cartesian space and {\bf b)} the perturber's interaction orbit (see text for explanation).\\
    {\bf c)} shows the final eccentricities of the bound particles, and
    {\bf d)} shows the final semi-major axes of the bound particles relative to the radii of their initial orbits. (\bf Image d) from Corrigendum.)\\
  }
  \label{fig:m01_5x5}
\end{figure*}

\begin{figure*}[h!t]
  \centering
  \begin{minipage}[t]{\hsize}
    \centering
    \begin{minipage}[t]{0.42\hsize}
      \vspace{0pt}
      \begin{subfigure}[t]{\textwidth}
        \begin{minipage}[t]{0.05\textwidth}
          \vspace{0pt}
          \caption{}\label{fig:m20_5x5_fates}
        \end{minipage}
        \hfill
        \begin{minipage}[t]{0.94\textwidth}
          \vspace{0pt}
          \includegraphics[width=\textwidth]{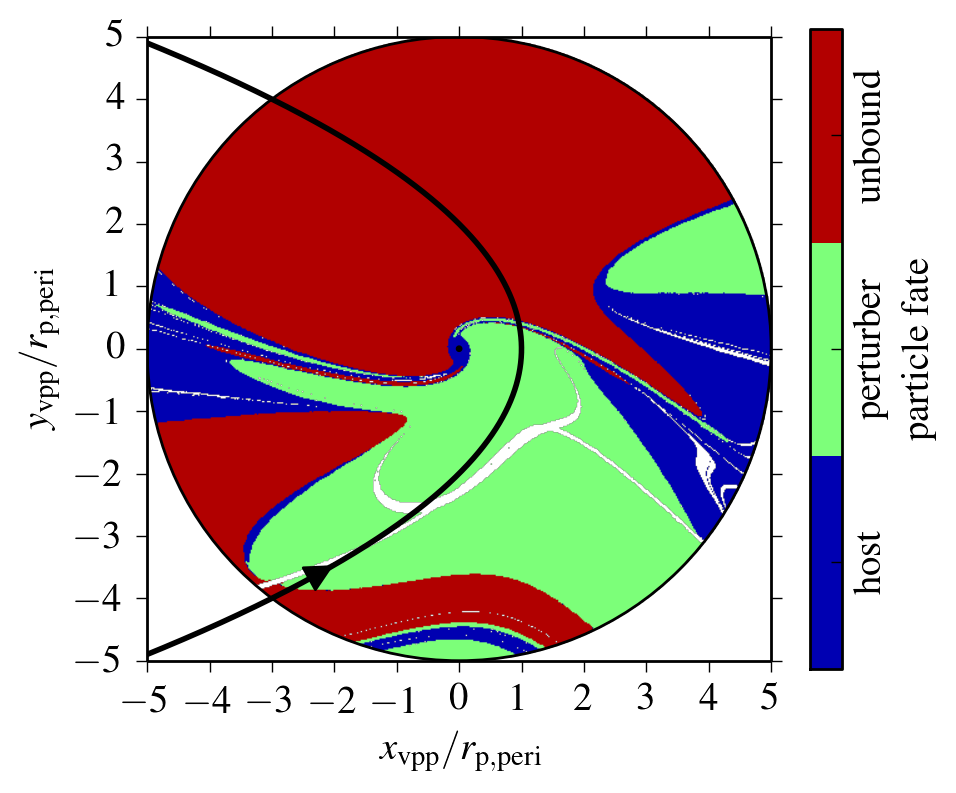}
        \end{minipage}
      \end{subfigure}
    \end{minipage}
    \hspace{1em}
    \begin{minipage}[t]{0.42\hsize}
      \vspace{0pt}
      \begin{subfigure}[t]{\textwidth}
        \begin{minipage}[t]{0.05\textwidth}
          \vspace{0pt}
          \caption{}\label{fig:m20_5x5_fates_io}
        \end{minipage}
        \hfill
        \begin{minipage}[t]{0.94\textwidth}
          \vspace{0pt}
          \includegraphics[width=\textwidth]{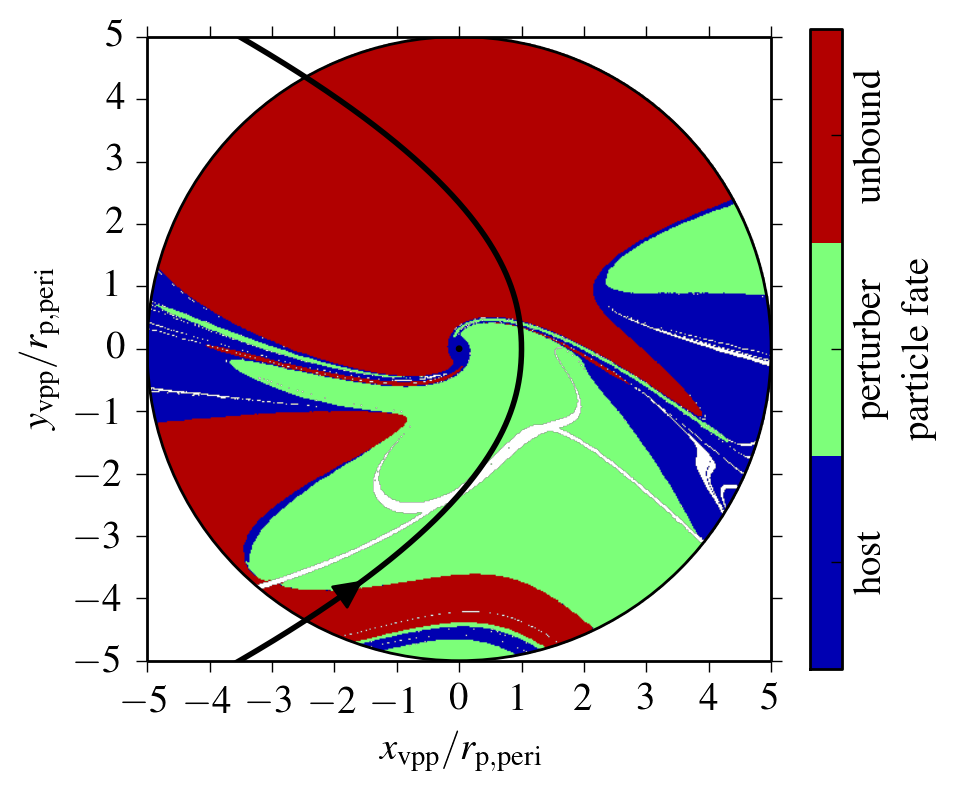}
        \end{minipage}
      \end{subfigure}
    \end{minipage}
  \end{minipage}
  \begin{minipage}[t]{\hsize}
    \centering
    \begin{minipage}[t]{0.42\hsize}
      \vspace{0pt}
      \begin{subfigure}[t]{\textwidth}
        \begin{minipage}[t]{0.05\textwidth}
          \vspace{0pt}
          \caption{}\label{fig:m20_5x5_ecc_bound}
        \end{minipage}
        \hfill
        \begin{minipage}[t]{0.94\textwidth}
          \vspace{0pt}
          \includegraphics[width=\textwidth]{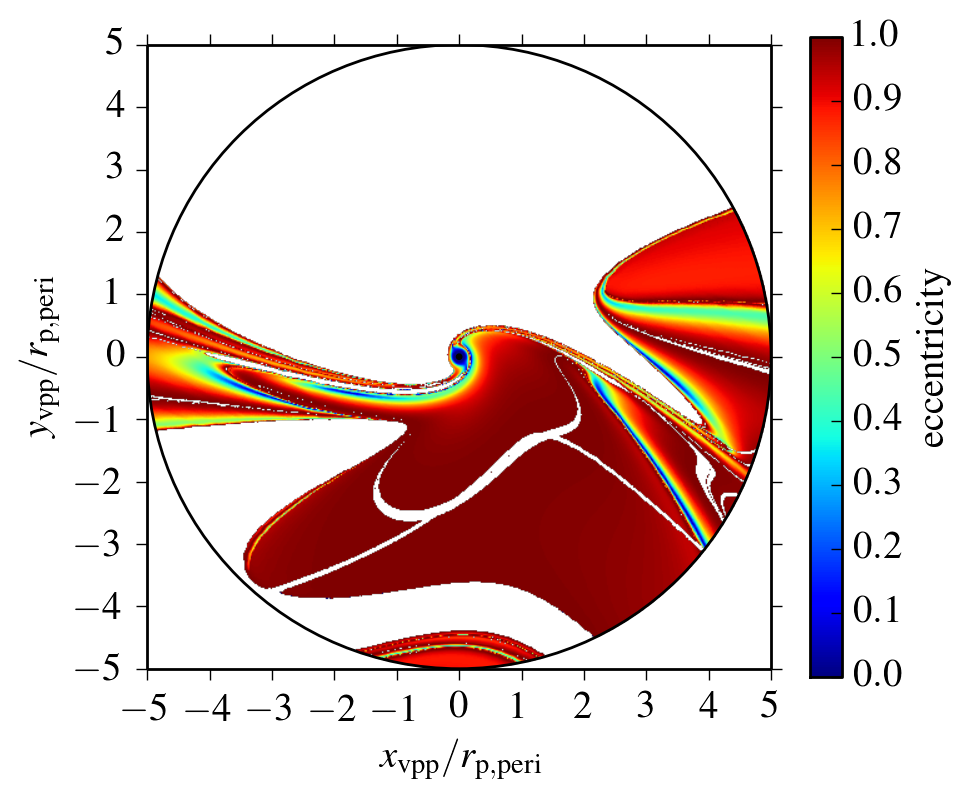}
        \end{minipage}
      \end{subfigure}
    \end{minipage}
    \hspace{1em}
    \begin{minipage}[t]{0.42\hsize}
      \vspace{0pt}
      \begin{subfigure}[t]{\textwidth}
        \begin{minipage}[t]{0.05\textwidth}
          \vspace{0pt}
          \caption{}\label{fig:m20_5x5_semimajor_bound}
        \end{minipage}
        \hfill
        \begin{minipage}[t]{0.94\textwidth}
          \vspace{0pt}
          \includegraphics[width=\textwidth]{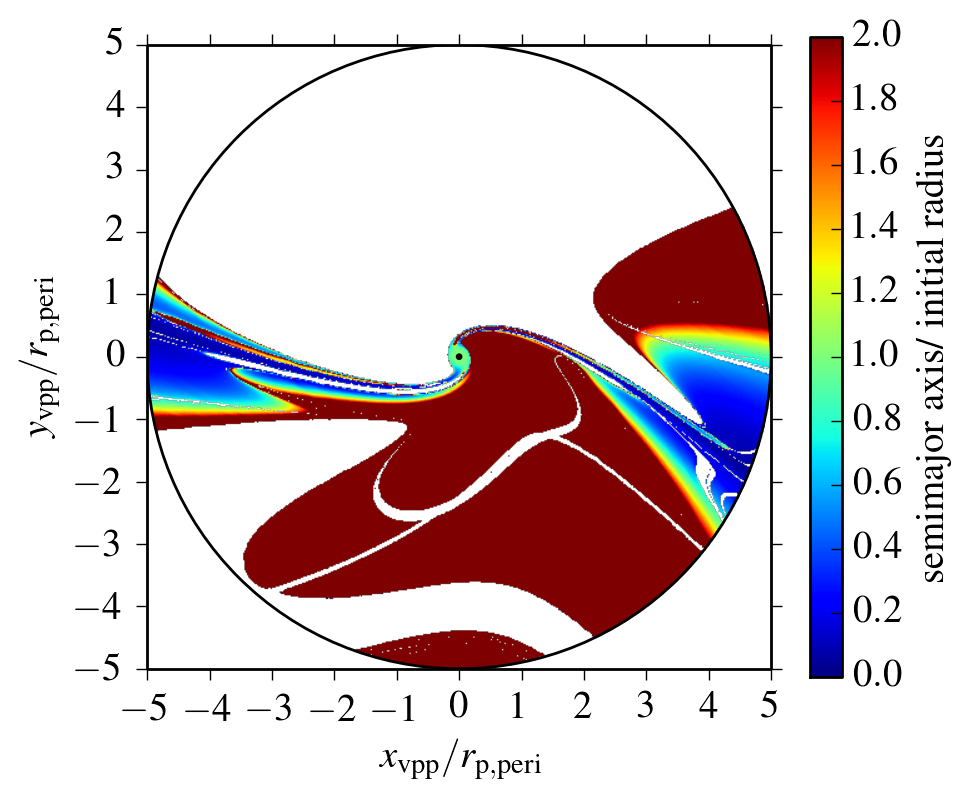}
        \end{minipage}
      \end{subfigure}
    \end{minipage}
  \end{minipage}
  \caption{As for Fig.~\ref{fig:m01_5x5}, but for a mass-ratio of $m = 20.0$. (\bf Image d) from Corrigendum.)}
  \label{fig:m20_5x5}
\end{figure*}

From our simulations, we determined the fates of the particle trajectories: whether the particles remain bound to the host,
are captured by the perturber, or become completely unbound.
Additionally, the final orbital elements of the particles were determined. They define the final orbits completely.
Depending on the particles
VPPs, these final properties can be presented as maps.
Because we want to present the new parametrisation of the initial conditions, along with some consequences here,
we only show some of the final properties for three sample mass-ratios.

\subsection{Low-mass perturbation}
\label{sec:low_mass}

Figure~\ref{fig:m01_5x5} shows the maps for the region with $r_{\mathrm{vpp}} /r_{\mathrm{p,peri}}  < 5$
for the mass-ratio $m = 0.1$.
This corresponds to the perturbation of a disc with initial radius of $r_{\mathrm{disc}} = 5\, r_{\mathrm{p,peri}} $,
or a perturbation with $r_{\mathrm{p,peri}}  = 0.2\, r_{\mathrm{disc}}$, respectively.
Initially, the particles were orbiting the host (indicated by the black dot at \coord{0}{0}) counter clockwise.
The perturber's pericentre is at \coord{0}{1}.

Figure~\ref{fig:m01_5x5_fates} shows the particles' fates:
particles with VPPs in the blue region remain bound to the host,
particles from the green region
\footnote{To improve the readability we use ``particles from ... region''
  synonymous for ``particles with VPPs in ... region''.
}
are captured by the perturber, and particles from the red region become unbound.
The black line in Fig.~\ref{fig:m01_5x5_fates} shows the orbit of the perturber --
the perturber came from the bottom edge of the image (as indicated by the black arrow), moved to \coord{1}{0}, and left across the top edge.
The white regions indicate particles for which the accuracy achieved with the used integration method is not sufficient,
and which are therefore not shown (see below for further explanation and Sect.~\ref{sec:discussion}).

It is often assumed that there is a simple radial dependence on whether particles remain bound or not.
Figure~\ref{fig:m01_5x5} shows that this is certainly not the case, but a much more complicated dependence exists.
It can be seen that only particles with VPPs in a narrow band along the approaching branch of the perturber's orbit
(black line in Fig.~\ref{fig:m01_5x5_fates}) become unbound.
On the side of the departing branch of the perturber's orbit, particles with VPPs from a broader region become unbound.
The only region from where particles are captured by the perturber is close to the perturber’s pericentre (green area around (1,0)).
Owing to the low perturber mass-ratio, the region from where particles become unbound is relatively small.

As can be seen in Fig.~\ref{fig:m01_5x5_fates}, the shape of the perturber orbit cannot be used to explain,
for example, the shape of the region from where particles become unbound.
The reason is that the perturber orbit shows all positions of the perturber for a time span
while the map shows the fates of particles which are at a certain position at a certain moment.
Even under the assumption that the particles are not influenced by the perturber at all,
particles that meet the perturber are not shown near the perturber orbit in the map.
When the perturber is at the coordinates ($r_{1}$, $\theta_{1}$) at the time $t_{1}$,
a particle being exactly at this position at that time is shown in the map at the position
\mbox{($r_{1}$, $\theta_{1} + \sqrt{\mu/r_{1}^3} \, (t_{\mathrm{peri}}  - t_{1})$)}.
These positions are highlighted for all $t_{1}$ with the black line in Fig.~\ref{fig:m01_5x5_fates_io}
and are referred to in the following as the perturber's ``interaction orbit''.

Figure~\ref{fig:m01_5x5_fates_io} shows that the interaction orbit matches the shape of the region from where particles become unbound much better than the real perturber orbit.
Comparing Figs.~\ref{fig:m01_5x5_fates} and~\ref{fig:m01_5x5_fates_io}, we can also see that part of the white dots in Fig.~\ref{fig:m01_5x5_fates} are very close to the interaction orbit.
This means that particles with VPPs in these regions are directly on the way of the perturber and approach the perturber very closely.
This results in integration errors, which is why these particles are excluded from the results.

Besides the fates of the particles, the shapes of their final orbits are important to describe the outcome of the perturbation process.
The shapes of the orbits are defined by their eccentricities and semi-major axes.
These two properties can also be transferred into energy and angular momentum to obtain another, often used, view on the perturbation process.

For the particles that are finally bound to the host or perturber,
the final eccentricities ($e < 1$) and semi-major axes relative to the initial radii of their orbits
are shown in Figs.~\ref{fig:m01_5x5_ecc_bound} and \ref{fig:m01_5x5_semimajor_bound}.
Since the unbound particles are of minor importance for most applications,
their final orbital elements can be found in Fig.~\ref{fig:5x5_unbound} in the appendix.

The final eccentricities of the particles that remain bound to the host
can be considered as nearly symmetrical to the $x$-axis (see Fig.~\ref{fig:m01_5x5_ecc_bound}).
Close to the region from where the particles become unbound (white region in Fig.~\ref{fig:m01_5x5_ecc_bound}),
the eccentricities are very high, up to $e = 1$.
With increasing distance to the white region, the eccentricities decrease.
Around the host, there is an approximately round area, where the particles are still on circular orbits, i.e. the
eccentricities are close to $0$.

Figure~\ref{fig:m01_5x5_semimajor_bound} shows the semi-major axes of the particles that are finally bound to the host or perturber relative to the radius of the initial particle orbit around the host.
{\bf The final semi-major axes have large values ($\gtrsim 2$) along the regions where the particles become unbound.
  With increasing distance from these regions, the values of the final semi-major axes decrease.
  This decrease with distance is steeper in the regions where the particles move ahead of the perturber (e.g. around \mbox{\coord{1}{-3}} and \mbox{\coord{1}{2}})
  and is shallower where the particles move behind the perturber (e.g. around \mbox{\coord{-1}{4}} and \mbox{\coord{3}{1}}).
  Along the negative x-axis the values of the final semi-major axes are close to the radii of the initial orbits (green region).
  Above the negative x-axis there is a larger, elliptical region where the final semi-major axes are also larger
  even though the region is far away from the region where the particles become unbound.
  Ahead of the departing branch of the perturber's interaction orbit there is a wider band where the final semi-major axes are much smaller
  than the radii of the particle's initial orbits.
  All captured particles have finally semi-major axes that are much smaller than the radii of their initial orbits,
  what is in agreement with the finding of \citet{2005ApJ...629..526P} that the captured particles are usually on very tight orbits.\footnote{\bf Text from Corrigendum.}
}

For the mass-ratio shown, we can generalise that,
as expected, the closer the VPP of a particle to the perturber's interaction orbit (see above),
the stronger it is influenced:
particles with VPPs very close to the interaction orbit become unbound,
particles with VPPs far away from the interaction orbit remain bound to the host with relatively low eccentricities.

Since the particles initially rotated counter-clockwise, the cleared area (red area in Fig.~\ref{fig:m01_5x5_fates})
mostly lags behind the interaction orbit.
Particles which move ahead
\footnote{Ahead and behind in the following mean ahead and behind in the direction of rotation.}
of the perturber are decelerated by its gravitational attraction.
These particles lose part of their energy and/or angular momentum,
and remain finally often on an eccentric orbit, bound more strongly to the host than before.
In contrast, particles which move behind the perturber are accelerated.
If the acceleration is strong enough they gain enough energy to reach escape velocity and become unbound.
The rear (in direction of rotation) borders of the red region in Fig.~\ref{fig:m01_5x5_fates}
therefore mark those particles that gain enough energy to exactly reach escape velocity.
The particles beyond these lines gain less energy and remain bound to the host on orbits of decreasing eccentricity
with increasing distance from the lines.
Since particles ahead of the interaction orbit are decelerated and remain bound to the host,
while particles behind are accelerated and may become unbound,
the perturber's interaction orbit describes approximately
the front border (in direction of rotation) of the cleared region.

\subsection{High-mass perturbation}
\label{sec:high_mass}

We now show the extreme case of a perturber with a relative mass of $ m = 20$.
As can be seen in Fig.~\ref{fig:m20_5x5}, the results differ considerably from the previous case
since most of the material from the investigated parameter space is now removed, either by becoming unbound
or by being captured by the perturber.
Only particles with VPPs in relatively small regions remain bound to the host (blue regions in Fig.~\ref{fig:m20_5x5_fates}).
These regions are mainly around the host and along the negative and positive $x$-axis.
Interestingly, only very few particles with VPPs between $r_{\mathrm{vpp}}  /r_{\mathrm{p,peri}}  \approx 0.3$ and $r_{\mathrm{vpp}}  /r_{\mathrm{p,peri}}  \approx 2$ remain bound to the host.
Most of the particles from the upper half of the parameter space become unbound,
from the lower half most particles are captured by the perturber.

Figure~\ref{fig:m20_5x5_fates_io} shows that the interaction orbit (see Sect.~\ref{sec:low_mass})
does not make much sense in this case.
This is a consequence of different dominating physical processes for different mass-ratios.

\begin{figure}[b!]
  \centering
  \includegraphics[width=0.6\hsize]{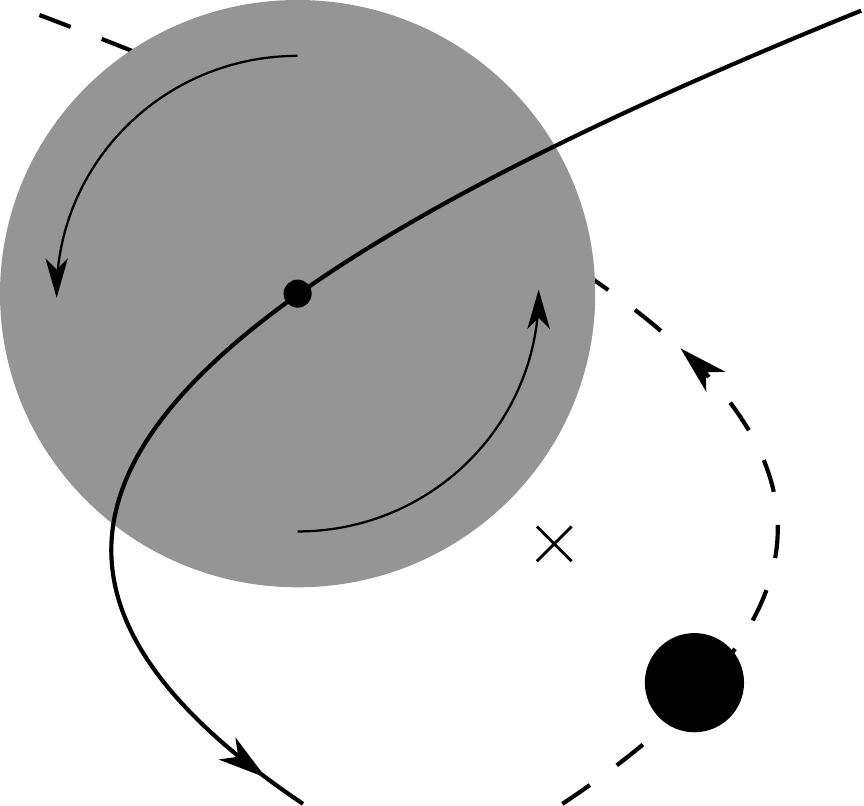}
  \caption{Relative orbits in the case of a high-mass perturber.
    The centre of mass is indicated by the cross.
    The solid line depicts the orbit of the host and the dashed line the orbit or the perturber.
    The grey circle denotes the low-mass material and the arrows show the rotation directions.}
  \label{fig:rel_orbit}
\end{figure}

In the case of a high-mass perturber ($m \gg 1$), the perturber rests close to the centre of mass,
while the host and the particles move on a wide orbit.
The distance between host and perturber, and the relative velocity of the host is thereby defined by the parabolic orbit.
In the case of a prograde encounter, the particles outside the host's orbit
(see also Fig.~\ref{fig:rel_orbit}) have a higher velocity relative to the perturber
owing to their superimposed rotation around the host.
Since the perturber's attraction is weaker outside the host's orbit, these particles become unbound.
This is the case for the particles in the upper half of your parameter space.
The removal of these particles can be described as centrifugal clearing.
The particles inside the host's orbit have lower velocities relative to the perturber
than necessary to overcome the perturber's gravitational attraction at their positions.
These particles mainly become captured by the perturber.

Just as the interaction orbit helped to explain the final properties of the particles in the low-mass perturbation case,
this process helps to explain them for the high-mass case.
Especially, it explains why the particles from the upper half of our parameter space are differently affected
during the perturbation than the particles from the lower half.

As in the low-mass case (see Sect.~\ref{sec:low_mass}), particles from an approximately circular region around the host finally have
relatively low eccentricities ($e \approx 0$, see Fig.~\ref{fig:m20_5x5_ecc_bound}).
In contrast, most of the particles that are captured by the perturber have very high eccentricities.

Nearly all particles that remain bound to the host finally have semi-major axes that are significantly smaller than the radii of
their initial orbits around the host (see Fig.~\ref{fig:m20_5x5_semimajor_bound}).
Only particles which almost become captured by the perturber or become unbound remain bound to the host with 
larger final semi-major axes (e.g. the red region at \mbox{$\approx$ \coord{3}{0.8}} to \coord{5}{0.8} in Fig.~\ref{fig:m20_5x5_semimajor_bound}).
{\bf The particles that are captured by the perturber have, in general, final semi-major axes which are much larger (factor $\gtrsim 2$)
  than the radii of their initial orbits around the host.\footnote{\bf Text from Corrigendum.}}

A signature of the centrifugal clearing can also be seen in
the final periapsides with the centre of mass of the unbound particles (see Fig.~\ref{fig:m20_5x5_periapsis_unbound} in the appendix).
The particles that are subject to the centrifugal clearing are ejected roughly in the direction
they were moving shortly before periastron passage, which is from right to left (see also Fig.~\ref{fig:rel_orbit}).
Their hyperbolic orbits basically originate from the positions where they were at that time.
Therefore, the periapsides of these particles can approximately be explained with the initial radii of their orbits around the host
plus the distance between host and perturber at the moment of ejection, which is $\approx 1$.
For example, a particle from (0,1) has an initial orbital radius of $1$ and a final periapsis of $\approx 2$.

\subsection{Equal-mass perturbation}
\label{sec:equal_mass}

\begin{figure*}[h!t!]
  \centering
  \begin{minipage}[t]{\hsize}
    \centering
    \begin{minipage}[t]{0.42\hsize}
      \vspace{0pt}
      \begin{subfigure}[t]{\textwidth}
        \begin{minipage}[t]{0.05\textwidth}
          \vspace{0pt}
          \caption{}\label{fig:m1_5x5_fates}
        \end{minipage}
        \hfill
        \begin{minipage}[t]{0.94\textwidth}
          \vspace{0pt}
          \includegraphics[width=\textwidth]{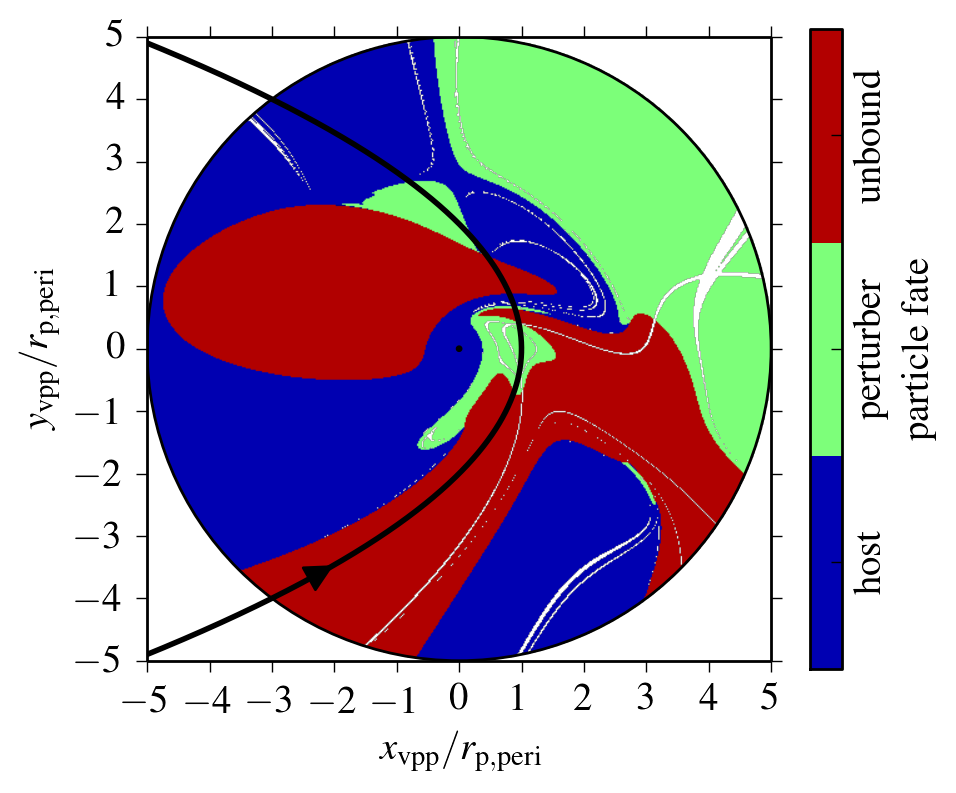}
        \end{minipage}
      \end{subfigure}
    \end{minipage}
    \hspace{1em}
    \begin{minipage}[t]{0.42\hsize}
      \vspace{0pt}
      \begin{subfigure}[t]{\textwidth}
        \begin{minipage}[t]{0.05\textwidth}
          \vspace{0pt}
          \caption{}\label{fig:m1_5x5_fates_io}
        \end{minipage}
        \hfill
        \begin{minipage}[t]{0.94\textwidth}
          \vspace{0pt}
          \includegraphics[width=\textwidth]{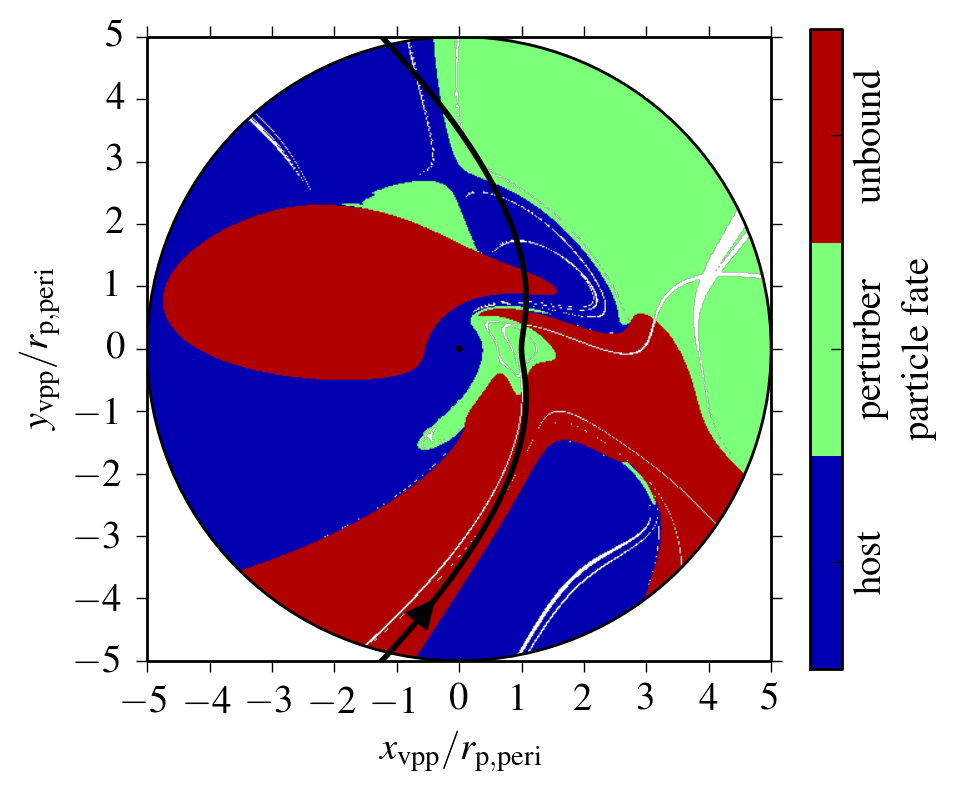}
        \end{minipage}
      \end{subfigure}
    \end{minipage}
  \end{minipage}
  \begin{minipage}[t]{\hsize}
    \centering
    \begin{minipage}[t]{0.42\hsize}
      \vspace{0pt}
      \begin{subfigure}[t]{\textwidth}
        \begin{minipage}[t]{0.05\textwidth}
          \vspace{0pt}
          \caption{}\label{fig:m1_5x5_ecc_bound}
        \end{minipage}
        \hfill
        \begin{minipage}[t]{0.94\textwidth}
          \vspace{0pt}
          \includegraphics[width=\textwidth]{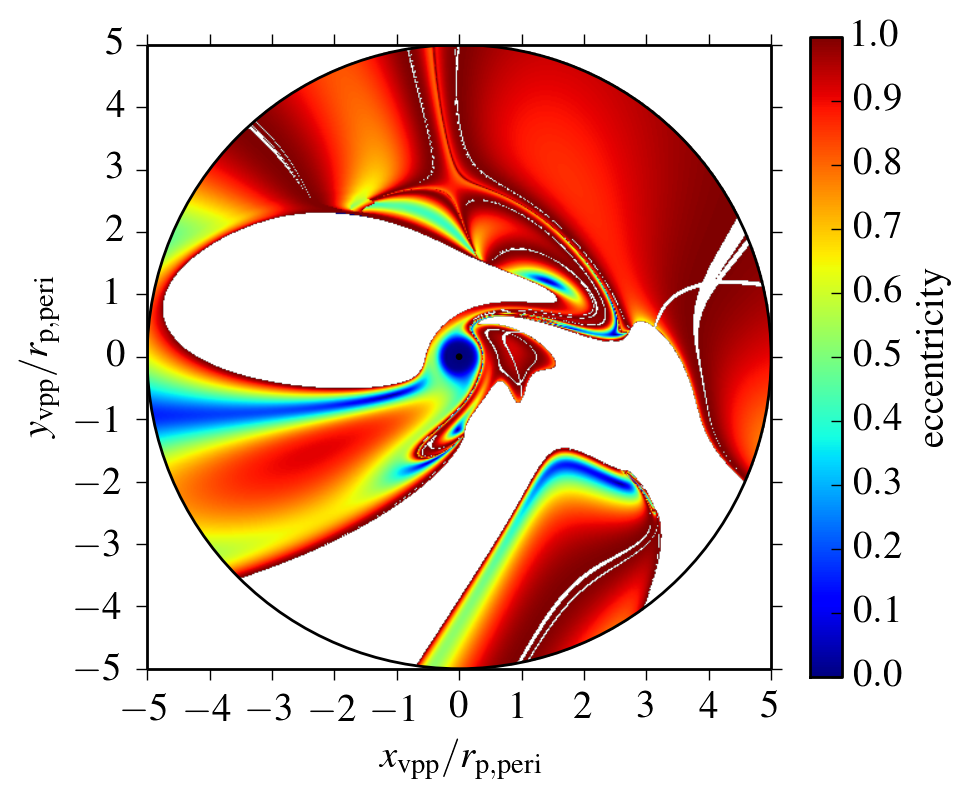}
        \end{minipage}
      \end{subfigure}
    \end{minipage}
    \hspace{1em}
    \begin{minipage}[t]{0.42\hsize}
      \vspace{0pt}
      \begin{subfigure}[t]{\textwidth}
        \begin{minipage}[t]{0.05\textwidth}
          \vspace{0pt}
          \caption{}\label{fig:m1_5x5_semimajor_bound}
        \end{minipage}
        \hfill
        \begin{minipage}[t]{0.94\textwidth}
          \vspace{0pt}
          \includegraphics[width=\textwidth]{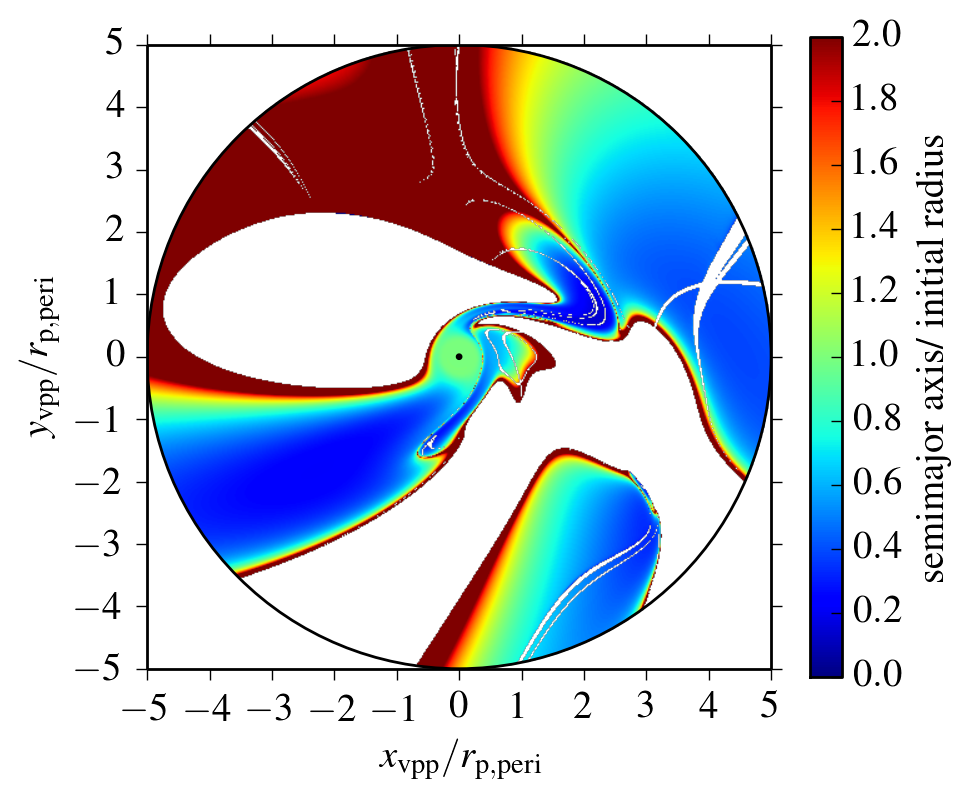}
        \end{minipage}
      \end{subfigure}
    \end{minipage}
  \end{minipage}
  \caption{As in Fig.~\ref{fig:m01_5x5}, but for a mass-ratio of $m = 1.0$. (\bf Image d) from Corrigendum.)}
  \label{fig:m1_5x5}
\end{figure*}

Comparing the results for the equal-mass case ($m = 1$) with the low-mass and the high-mass case,
one can see (in Fig.~\ref{fig:m1_5x5}) that the equal-mass case shows features of both other cases.
As in the low-mass case, particles from a band along the approaching branch of the perturber orbit become unbound.
Owing to the higher perturber mass the band is broader.
The rear border of this area again depicts the region where particles exactly gain escape velocity and
the front border can also be described approximately with the perturber's interaction orbit (see Fig.~\ref{fig:m1_5x5_fates_io}).

The red, drop-shaped region around \coord{-2}{1} in Fig.~\ref{fig:m1_5x5_fates} is centrifugally cleared as in the high-mass case.
Because of the lower perturber mass, the region is smaller.
All other features, especially the red and green regions in the right half of Fig.~\ref{fig:m1_5x5_fates},
are caused by repeated interactions of the particles with host and perturber.

In contrast to the low-mass case, the final eccentricities and semi-major axes of the particles (see Figs.~\ref{fig:m1_5x5_ecc_bound} and~\ref{fig:m1_5x5_semimajor_bound})
do not show any symmetry.
However, the finding that close to the cleared regions, the final eccentricities are very high ($e \approx 1$), still holds.
The nearly unperturbed region around the host is smaller than in the low-mass case and bigger than in the high-mass case.

Most particles that remain bound to the host or are captured by the perturber have final semi-major axes, which are smaller than
the radii of their initial orbits.
{\bf In contrast, almost all particles with VPPs in the second quadrant (see Fig.~\ref{fig:m1_5x5_semimajor_bound}) have finally semi-major axes that are more than twice   as large as the radii of their initial orbits.
  This region is a continuation of the region where the particles are centrifugally removed.
  Here, the energy gain is just not enough to become unbound.\footnote{\bf Text from Corrigendum.}}

Although an encounter between stars of equal mass was often taken as a standard case to investigate,
for example, the effect on disc properties \citep[e.g.][]{1993MNRAS.261..190C,1996MNRAS.278..303H},
the fate and final properties of the particles are not easy to explain.
This is because this case is rather a complicated mixture of low- and high-mass encounters.

\subsection{Final orbital elements of the bound particles}

\begin{figure}[t!]
  \centering
  \begin{minipage}[t]{\hsize}
    \vspace{0pt}
    \begin{subfigure}[t]{\textwidth}
      \begin{minipage}[t]{0.05\textwidth}
        \vspace{0pt}
        \caption{}\label{fig:ecc_0.1}
      \end{minipage}
      \hfill
      \begin{minipage}[t]{0.94\textwidth}
        \vspace{0pt}
        \includegraphics[width=\hsize]{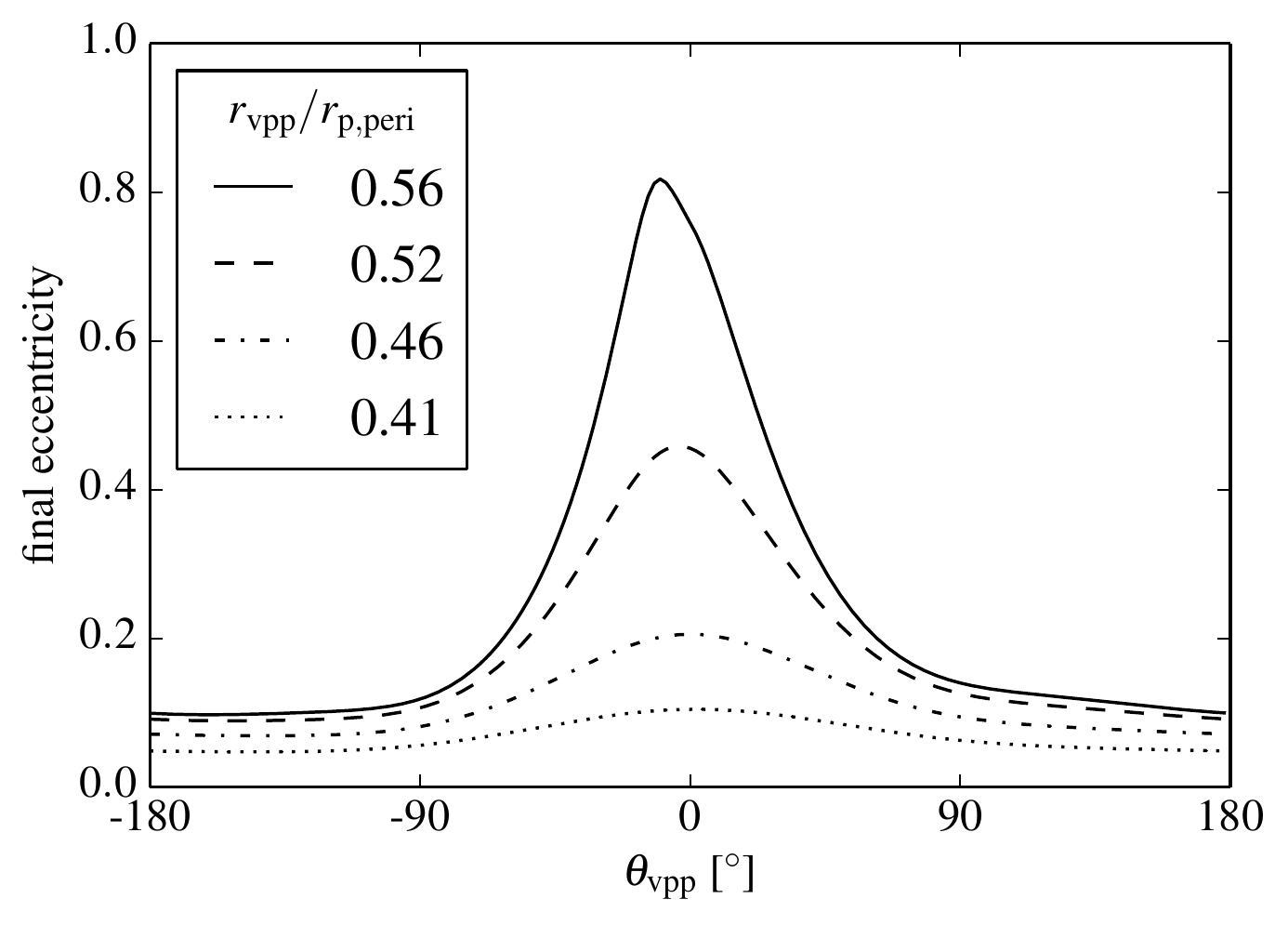}
      \end{minipage}
    \end{subfigure}
  \end{minipage}
  \hspace{1em}
  \begin{minipage}[t]{\hsize}
    \vspace{0pt}
    \begin{subfigure}[t]{\textwidth}
      \begin{minipage}[t]{0.05\textwidth}
        \vspace{0pt}
        \caption{}\label{fig:ecc_1.0}
      \end{minipage}
      \hfill
      \begin{minipage}[t]{0.94\textwidth}
        \vspace{0pt}
        \includegraphics[width=\textwidth]{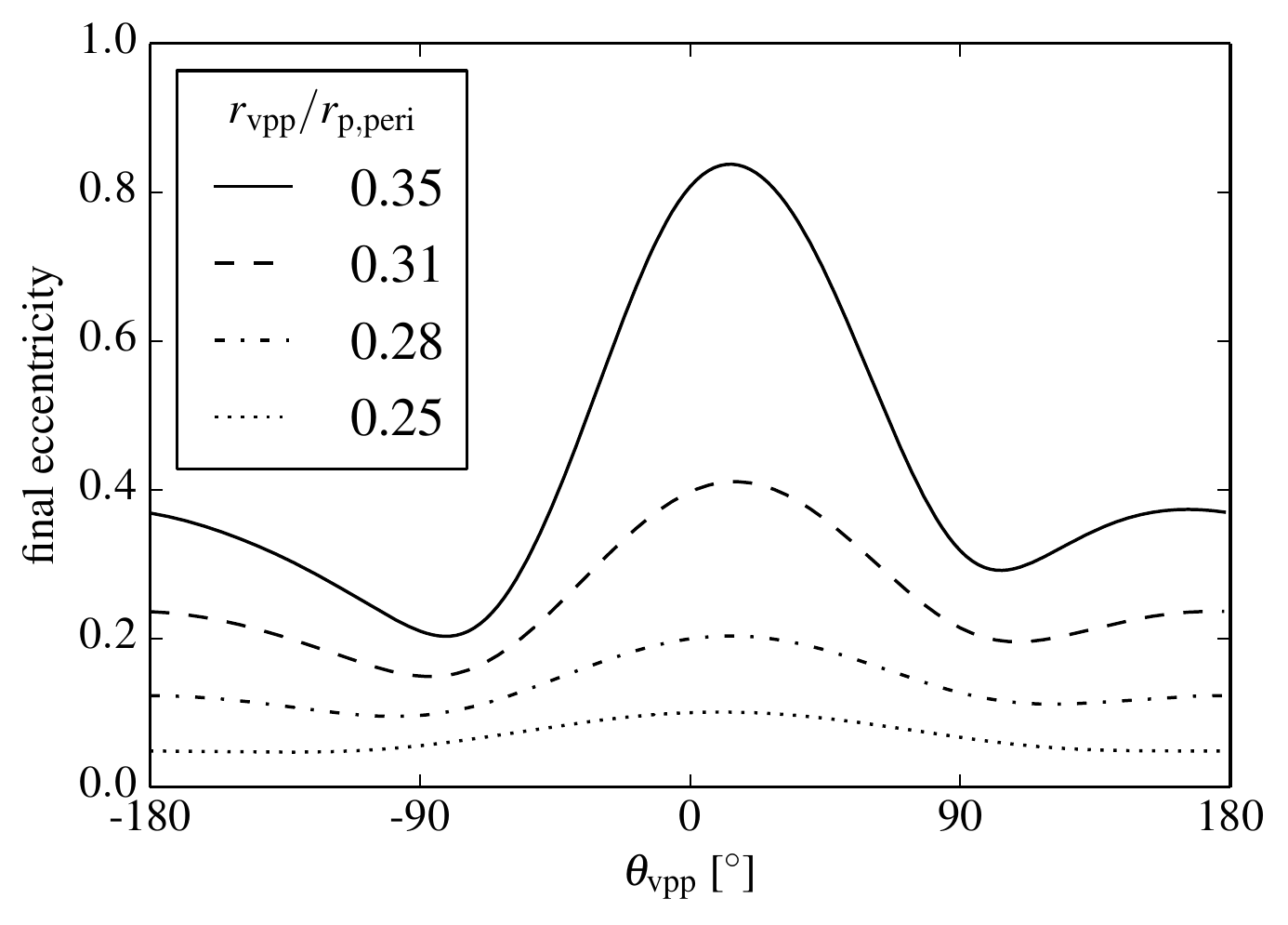}
      \end{minipage}
    \end{subfigure}
  \end{minipage}
  \hspace{1em}
  \begin{minipage}[t]{\hsize}
    \vspace{0pt}
    \begin{subfigure}[t]{\textwidth}
      \begin{minipage}[t]{0.05\textwidth}
        \vspace{0pt}
        \caption{}\label{fig:ecc_20.0}
      \end{minipage}
      \hfill
      \begin{minipage}[t]{0.94\textwidth}
        \vspace{0pt}
        \includegraphics[width=\textwidth]{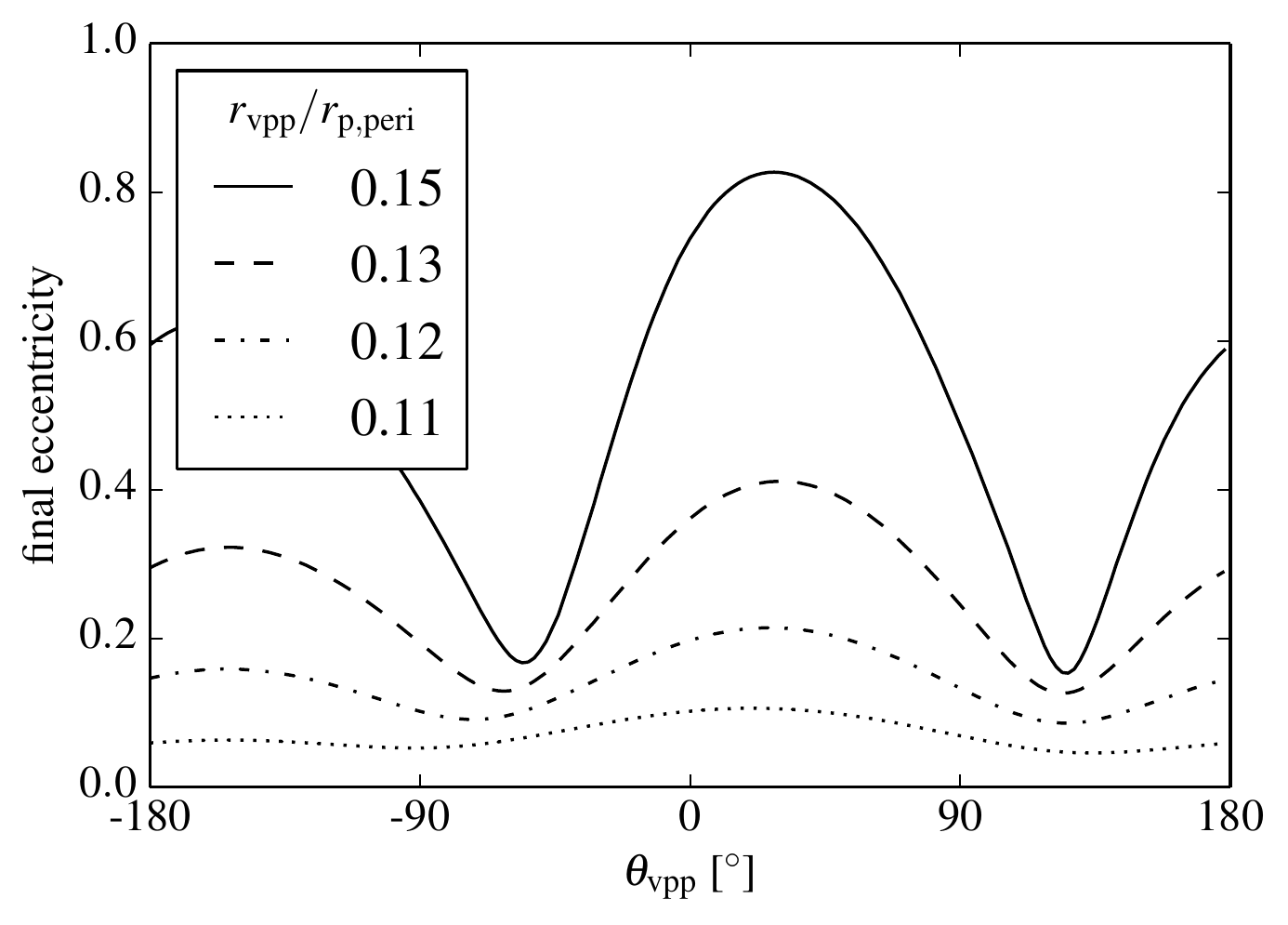}
      \end{minipage}
    \end{subfigure}
  \end{minipage}
  \caption{Final eccentricities of particles in the central region depending on the VPPs
    for the mass-ratios \mbox{$m = 0.1$ ({\bf a})}, \mbox{$m = 1.0$ ({\bf b})}, and \mbox{$m = 20.0$ ({\bf c})}.
  }
  \label{fig:functions}
\end{figure}

In the Figs.~\ref{fig:m01_5x5}, \ref{fig:m20_5x5}, and \ref{fig:m1_5x5}, it can be seen
that there are large regions in the investigated parameter space where particles have the same fate.
Furthermore, similar initial conditions within each of these regions usually result in similar final orbits,
indicated by the smooth colour gradients in the images with the final orbital elements of the particles
(e.g. Figs.~\ref{fig:m01_5x5_ecc_bound} and \ref{fig:m01_5x5_semimajor_bound}).
This knowledge can be used to interpolate the final orbital elements for initial conditions between the numerically integrated points.

For the inner parts of the investigated parameter space, where all particles remain bound to the host,
the final orbital elements of the particles follow relatively smooth functions of the radius and the angle, $\theta_{\mathrm{vpp}} $.
As an example, in Fig.~\ref{fig:functions}, the final eccentricities are shown for some sample radii ($r_{\mathrm{vpp}} /r_{\mathrm{p,peri}} $) as a function of $\theta_{\mathrm{vpp}} $.
In contrast to the data for Figs~\ref{fig:m01_5x5}, \ref{fig:m20_5x5}, and \ref{fig:m1_5x5},
these data were obtained from simulations where the parameter space was sampled in polar coordinates.
The data are not smoothed.

For a fixed $\theta_{\mathrm{vpp}} $, the final eccentricities are higher for larger initial radii, as expected.
Around $\theta_{\mathrm{vpp}}  \approx 0$, the final eccentricities are
increased relative to the values outside $ -90 \lesssim \theta_{\mathrm{vpp}}  \lesssim 90$.
The values seem to be a simple function of $\theta_{\mathrm{vpp}} $ and $r_{\mathrm{init}} /r_{\mathrm{p,peri}} $.
But the derivation of this dependency lies beyond the scope of this paper.

\section{Discussion}
\label{sec:discussion}

We have presented a method to obtain general numerical solutions for the perturbation of low-mass objects on circular Keplerian orbits around a massive object
by the fly-by of another massive object.
As for all numerical solutions, the quality of the results depends on the chosen spacial and temporal resolution.

The temporal resolution is limited by the used integrator.
Here, the LSODA integrator is a well-established integrator with error-controlled time-step size.
As the maximum relative error of the integrated values, the built-in default of $\approx 1.5 \cdot 10^{-8}$ was used.
Since the trajectory of each particle was integrated individually,
the obtained trajectories for most particles are much more precise than, for example, those of \citet{2014A&A...565A.130B}.
Only for some particles that approach one of the massive objects so close that very small time steps are required,
the accuracy was deemed insufficient.
Therefore, particles that approach one of the two massive objects closer than $0.0001 \cdot r_{\mathrm{p,peri}} $ have been excluded from the shown images (see Sect.~\ref{sec:low_mass}).

Other potentially critical factors for the accuracy of the results are the duration of the simulation before ($t_{\mathrm{peri}}  - t_{\mathrm{init}} $)
and after ($t_{\mathrm{end}}  - t_{\mathrm{peri}} $) pericentre passage of the perturber.
By varying the initial force influence of the perturber (see also Sect.~\ref{sec:numerics})
and hence $t_{\mathrm{peri}}  - t_{\mathrm{init}} $, we investigated the influence of the starting time on the results.
We found an initial relative force influence of $F_{\mathrm{p}}/F_{\mathrm{h}} \lesssim 10^{-4}$ to be sufficient for our purpose.

The results also converge for $t_{\mathrm{end}}  - t_{\mathrm{peri}}  \to \infty$.
By integrating the particle trajectories individually until the orbital elements do not change more than $1$~\% during a certain time span
(see also Sect.~\ref{sec:numerics}), we ensure that our results are also converged with respect to $t_{\mathrm{end}} $.

For cases where the orbits of the low-mass particles are not exclusively determined by the masses of host and perturber,
e.g. due to self-gravity and / or viscosity, our results can not directly be applied.
Since, in these cases, the initial angular velocity of the particles
no longer only depends on the distance to and the mass of the host,
the particle's VPPs cannot be determined analytically.
However, as long as additional forces are small, compared to the gravitational force of the perturber around pericentre,
it is still possible to make some predictions since the physical processes will be similar.

For example, for self-gravitating, non-viscous discs, it can be expected
that a low-mass perturber also clears only a narrow band along its orbit, depending on the range of its force.
The eccentricities of the remaining particles will also decline with increasing distance to the perturber's interaction orbit.
Similarly, even self-gravity and viscosity can only weaken but not prevent the centrifugal clearing of the upper half
of our parameter space in the case of a high-mass perturber (see also Sect.~\ref{sec:high_mass}).
Especially viscosity does not provide a strong enough additional attracting force.
The repulsing effect of viscosity has also no impact here since the material is removed in a direction without other material.

In the case of viscous discs, the gradual exchange of energy and angular momentum leads to processes in the disc
like accretion and disc-spreading, even without a perturbation.
Therefore, predictions for viscous discs from our results may only be possible
for a period after the perturbation that is shorter than the viscous timescale.
Depending on the strength of viscosity, during this period the orbital elements of the material may deviate considerably from our results
since the influence of the perturber is not yet negligible and the particles have not yet settled into stable orbits.
After this short period, the viscous evolution leads to further change in the orbital elements, which makes a comparison difficult.
Additionally, predictions from our results may only be possible for material
that does not move through regions with higher density during and after the perturbation.

\section{Application to planetary systems and non-viscous discs}
\label{sec:application}

\begin{figure}[t!]
  \centering
  \begin{minipage}[t]{\hsize}
    \vspace{0pt}
    \begin{subfigure}[t]{\textwidth}
      \begin{minipage}[t]{0.05\textwidth}
        \vspace{0pt}
        \caption{}\label{fig:app_disc_1}
      \end{minipage}
      \hfill
      \begin{minipage}[t]{0.94\textwidth}
        \vspace{0pt}
        \centering
        \includegraphics[width=0.8\textwidth]{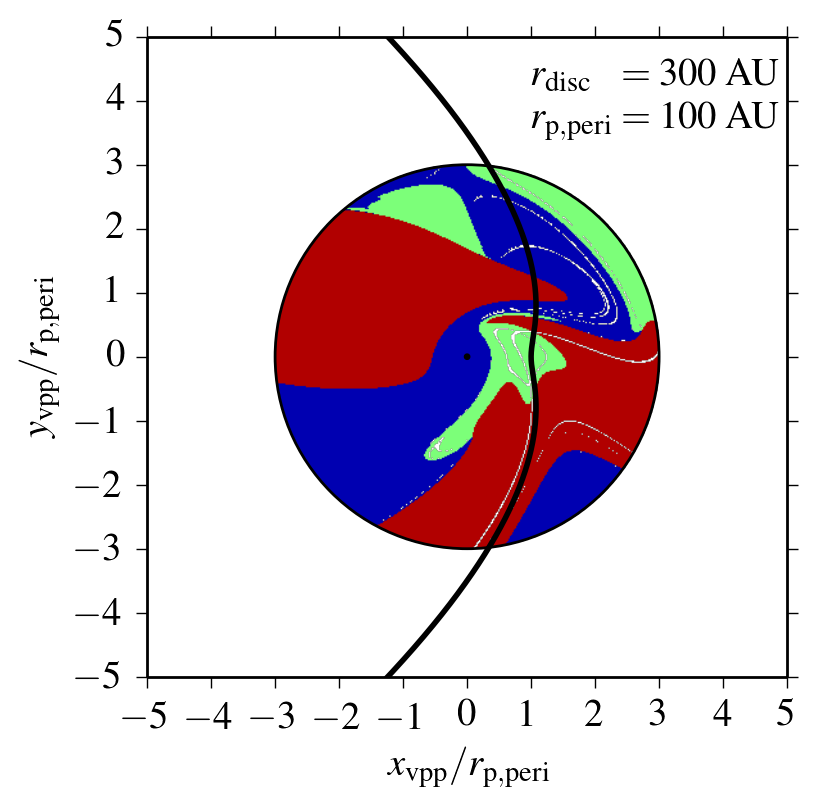}
      \end{minipage}
    \end{subfigure}
  \end{minipage}
  \hspace{1em}
  \begin{minipage}[t]{\hsize}
    \vspace{0pt}
    \begin{subfigure}[t]{\textwidth}
      \begin{minipage}[t]{0.05\textwidth}
        \vspace{0pt}
        \caption{}\label{fig:app_disc_2}
      \end{minipage}
      \hfill
      \begin{minipage}[t]{0.94\textwidth}
        \vspace{0pt}
        \centering
        \includegraphics[width=0.8\textwidth]{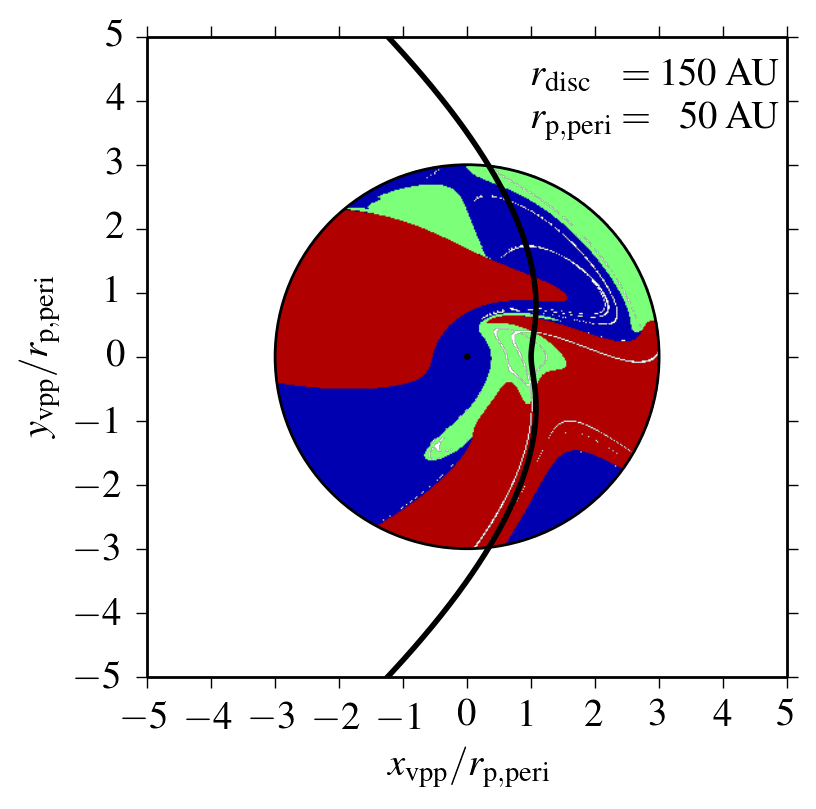}
      \end{minipage}
    \end{subfigure}
  \end{minipage}
  \hspace{1em}
  \begin{minipage}[t]{\hsize}
    \vspace{0pt}
    \begin{subfigure}[t]{\textwidth}
      \begin{minipage}[t]{0.05\textwidth}
        \vspace{0pt}
        \caption{}\label{fig:app_disc_3}
      \end{minipage}
      \hfill
      \begin{minipage}[t]{0.94\textwidth}
        \vspace{0pt}
        \centering
        \includegraphics[width=0.8\textwidth]{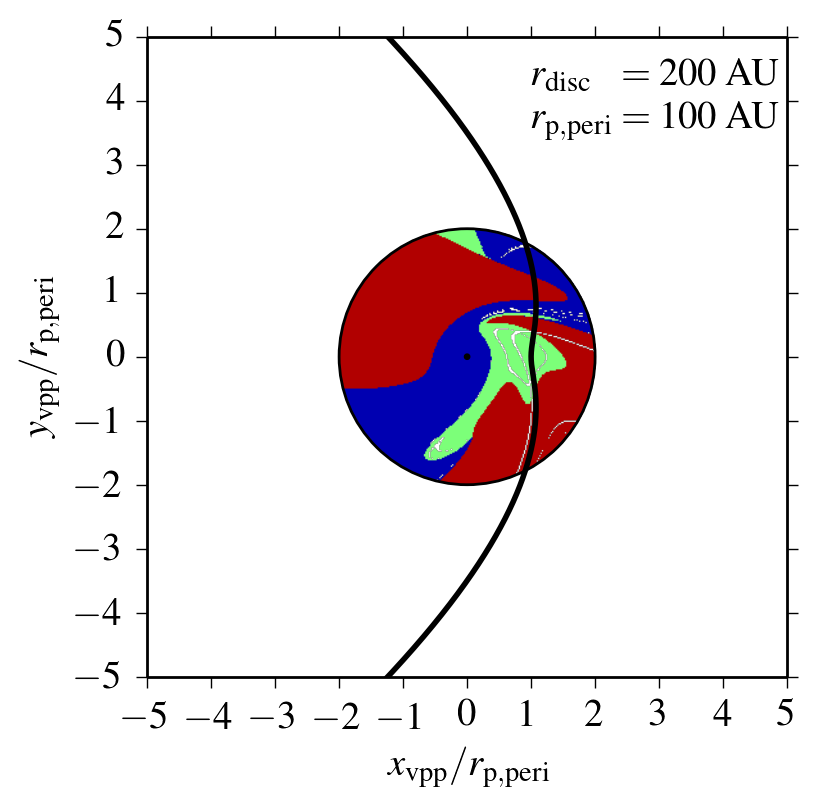}
      \end{minipage}
    \end{subfigure}
  \end{minipage}
  \caption{Sample application of the particle fates for non-viscous,
    low-mass discs with different initial sizes perturbed by a perturber with a mass-ratio of $m = 1.0$ and
    different pericentre distances.
    The scale is the same as in Fig.~\ref{fig:m1_5x5} for better comparison. For full description, see text.}
  \label{fig:application}
\end{figure}

The results of the restricted three-body approximation presented here are directly applicable to stars surrounded by low-mass discs,
either protoplanetary or debris, or planetary systems that are disturbed by the fly-by of a second star.
However, this approach is only valid if viscous forces can be neglected.
Furthermore, the matter has to be on circular orbits before the perturbation.

For the application to discs, we note
that the presented datasets contain information what happens to disc material for all encounters from 
penetrating (with \mbox{$r_{\mathrm{p,peri}}  = 0.2 \,r_{\mathrm{disc}} $} $\Leftrightarrow r_{\mathrm{disc}} /r_{\mathrm{p,peri}}  = 5$ ) up to very distant encounters.
As already pointed out by \citet{1972ApJ...178..623T}, the results for different initial disc sizes are contained in our dataset as subsets.
When investigating the effect of an encounter on a disc with a certain size, only the radial subset
which corresponds to the relative size of the respective disc is considered.

If, for example, a disc with an initial outer radius of $300$~AU is perturbed by a star with a periastron of
$100$~AU, only the parameter space up to
\mbox{$r_{\mathrm{init}} /r_{\mathrm{p,peri}}  = 300~\mathrm{AU}/100~\mathrm{AU} = 3$} would be considered.
For the equal mass case, this is illustrated in Fig.~\ref{fig:app_disc_1},
which shows only the inner area of Fig.~\ref{fig:m1_5x5_fates}.
Figure \ref{fig:app_disc_2} shows the same results for a disc with an initial outer radius of $150$~AU
perturbed by a star with a periastron of $50$~AU.
Since the whole problem scales with size, the result is exactly the same as for the previous case.
Figure \ref{fig:app_disc_3} shows the results for a disc with an initial outer radius of $200$~AU
perturbed by a star with a periastron of $100$~AU.
As can be seen, the result is the inner part of Fig.~\ref{fig:app_disc_1}.

The above example also illustrates that our method of representing the results of a perturbation is independent of the initial disc size as a parameter.
Previous investigations often performed simulations of perturbations of discs with different initial sizes \citep[e.g.][]{2005A&A...437..967P}.
But owing to the inherent geometrical scaling of the problem (see also Sect.~\ref{sec:method}), the obtained information was redundant.
Normalising the initial conditions with the perturber's pericentre distance, as already performed by, for example, \citet{1993MNRAS.261..190C}, removes this redundancy.
For the simulations for a parameter study, this is a reduction by one free parameter -- the initial disc size.
When investigating the effect of an encounter on a certain disc, this parameter must be introduced again.
But this is much faster than performing one simulation for each disc size.

Our method uses test particles to describe the effect of a fly-by on objects with VPPs in a given area.
How much matter resides initially in this area depends on the mass distribution within the disc.
To investigate the change in a physical quantity, like mass, angular momentum, or energy, one therefore needs
to attribute masses to the particles.

Furthermore, a real physical disc also has a vertical \mbox{mass / particle} distribution with a certain scale height,
which has to be considered.
The material, which initially moves on orbits that are inclined relative to the plane of the perturber orbit,
is affected differently than in the coplanar case we present here.
The resulting disc would then be a combination of the results for the coplanar case and the respective inclined cases.
Typical scale heights of these types of discs correspond to orbital inclinations $\lesssim 10^{\circ}$.
Tests have shown that for these inclinations the influence is of minor importance.

\begin{figure}[t!]
  \centering
  \includegraphics[width=\hsize]{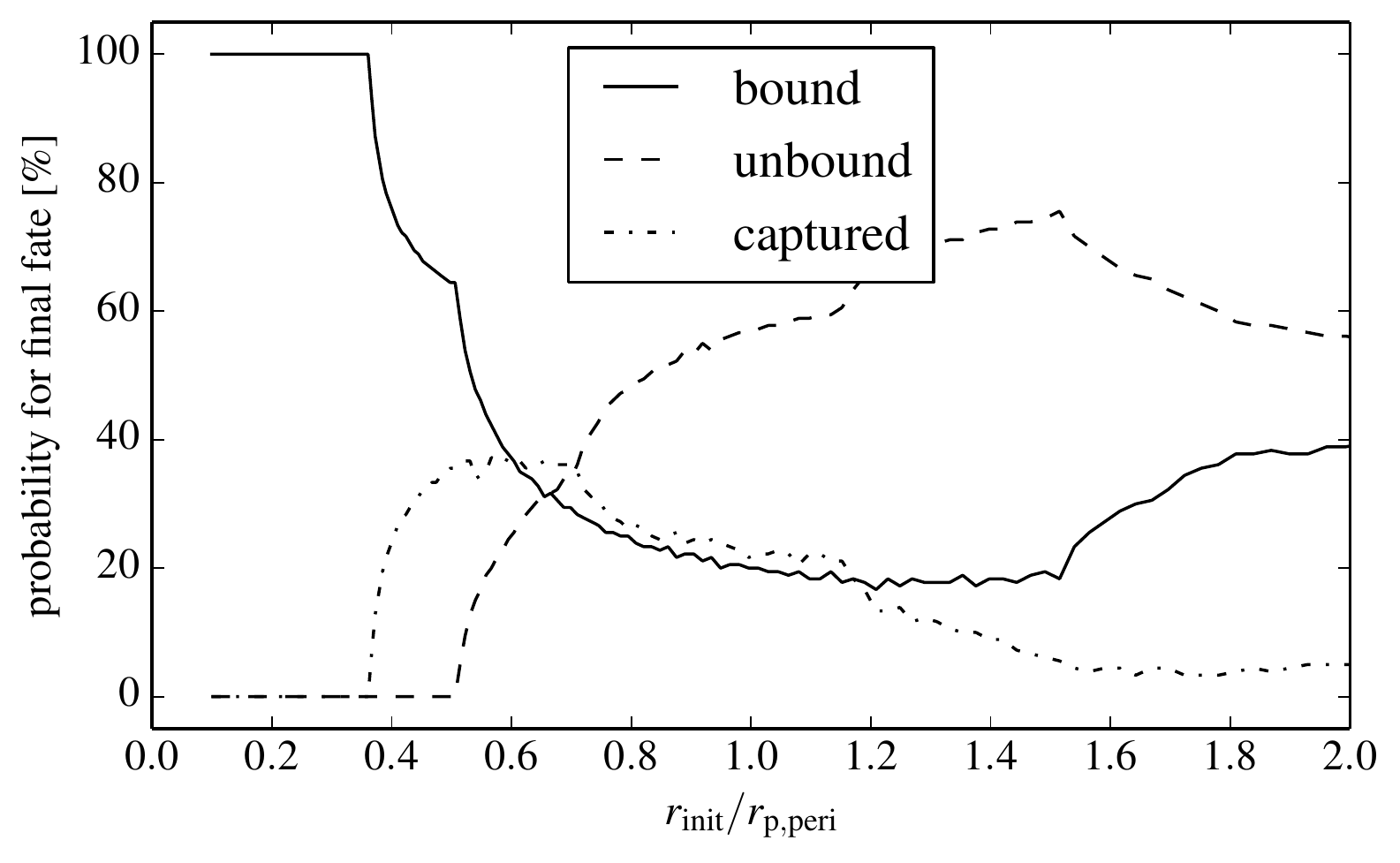}
  \caption{Probabilities for certain fates depending on the relative radius of a particle's initial orbit
    ($r_{\mathrm{init}} /r_{\mathrm{p,peri}} $) for the mass-ratio \mbox{$m = 1.0$}. See text for full description.
  }
  \label{fig:probabilities}
\end{figure}

In contrast to the application to discs, the post-encounter properties of a planet can be directly obtained from the maps
by scaling the initial radius of the planetary orbit to the perturber's periastron.
For example, the final properties of a planet initially on an orbit with $r=100$~AU around a star of mass $M_{\mathrm{h}}  = 1$~\mbox{M$_{\odot}$} 
perturbed by another star with mass $M_{\mathrm{p}}  = 1$~\mbox{M$_{\odot}$}  and a periastron of $r_{\mathrm{p,peri}}  = 200$~AU can be found in the maps at
$r_{\mathrm{vpp}} /r_{\mathrm{p,peri}}  = 0.5$.
Since, in practice, $\theta_{\mathrm{vpp}} $ may be unknown,
it might be sufficient to obtain the probability for a certain result depending on $r_{\mathrm{vpp}} /r_{\mathrm{p,peri}} $.
This method was already applied by \cite{1993MNRAS.261..190C} (see their Fig.~3.), but our
results are higher resolved and show less statistical scatter (see Fig.~\ref{fig:probabilities}).
Similarly to the data for Fig.~\ref{fig:functions}, the data for Fig.~\ref{fig:probabilities}
were obtained from simulations where the parameter space was sampled in polar coordinates.

\section{Summary}
\label{sec:summary}

In this work, we presented an efficient way to generalise the results of a perturbation
of low-mass bodies initially on circular Keplerian orbits around a massive object by the fly-by of another massive object.
In contrast to previous investigations, our new parametrisation maintains not only the radial resolution for 
the particles, but also the angular resolution.
By also exploiting the other known scaling relations of the problem,
the initial conditions for the low-mass bodies for one specific perturber orbit can now be specified with just
two parameters.
We have created maps that relate
the orbital properties of the low-mass bodies after the perturbation to this two dimensional initial condition space.
These maps enable us to see the correlation between the particle's initial position and its final properties.
This provides a better understanding of the underlying dynamics.

Maps of the particle fates (whether they remain bound to the host, are captured by the perturber, or become unbound)
and some final orbital elements of perturbed particle orbits (i.e. the final eccentricities and semi-major axes)
were presented for three sample mass-ratios.
With these maps, a detailed comparison of the perturbations with different mass-ratios is possible.
The maps show that a low-mass perturber ($M_{\mathrm{p}} /M_{\mathrm{h}}  < 1$) basically clears a narrow band along its orbit through the parameter space.
{\bf The closer to the perturber orbit, the higher the final eccentricities and the larger the semi-major axes
  of the particles that remain bound to the host.\footnote{\bf Text from Corrigendum.}}
For a high-mass perturber ($M_{\mathrm{p}} /M_{\mathrm{h}}  > 1$), the effect is much more violent -- only material from small regions remains finally bound.
In this case, the removal of particles from the host is a consequence of the centrifugal force acting on the particles
during the host's pericentre passage on its orbit around the heavier perturber.
The equal-mass case ($M_{\mathrm{p}} /M_{\mathrm{h}}  = 1$) is a mixture of the low-mass and the high-mass case.

Regions in the presented maps, where particles have the same fate, show smooth gradients in the final orbital elements.
As a consequence, the presented numerical results are not only valid for exactly the integrated initial conditions,
but can be generalised for the full presented parameter space by interpolation.

The results are directly applicable to astrophysical processes like the perturbation of planetary orbits
or protoplanetary discs by passing-by stars without the need for additional simulations.
The fates of planets in an encounter and the final orbital elements can directly be obtained from the maps
by scaling the initial planetary orbit with the perturber's periastron distance.
When applied to discs, our results are independent of the initial disc size, in contrast to most previous works.
The result of the perturbation of a disc with finite size can be obtained from our maps by considering only the
respective radial subset.
For the simulations of a parameter study in this context, this reduction by one parameter reduces the numerical effort significantly.

Even in the case of self-gravitating and viscous discs, the results may be applicable to some extent.
Material that is removed from the disc without moving through denser regions, as in the case of a high-mass perturber,
will most likely behave in a similar way.
By way of contrast, in the case of crossing particle trajectories, viscosity will change the results considerably.

\vspace{1em}
{\tiny {\bf Acknowledgements}: We thank the anonymous referee for very useful comments that helped to improve this paper significantly.
  AB also wants to thank A.~Bhandare, S.~Brackertz, T.~Bardenheuer, M.~Engler, C.~Korntreff, S.~Libi, K.-H.~Rehren, M.~Sonntag, A.~Wolff, and M.~Xiang-Gruess for fruitful discussions.
  Additionally, AB is indebted to his family, without whose support this work might not have been finished.}

\bibpunct{(}{)}{;}{a}{}{,} 

\bibliographystyle{aa}

\Online

\begin{appendix}

\section{Properties of the unbound particles}

In this section, we show the results for the particles that become unbound during the perturbation.
Astrophysical investigations of the perturbation of protoplanetary discs or planetary systems usually focus on matter
that remains bound to the host or is captured by the perturber. Therefore, these results are probably of minor importance.
However, the final orbital elements of the unbound particles support our explanation for the two different processes
for low- and high-mass perturbations.

For the unbound particles, we show the orbital elements with the centre of mass of host and perturber.
The unbound particles are on hyperbolic orbits with the centre of mass and have, therefore, negative semi-major axes.
Because these are difficult to interpret, we show instead the periapsis distances
\footnote{
  As the orbital elements are calculated from the final positions of the particles relative to the centre of mass
  and the centre of mass moves relative to host and perturber,
  it is difficult to transfer this information into the VPP system, which is host-centred.
  A small periapsis distance does not mean that a particle's hyperbolic trajectory originates from close to the host.
  It just means that this trajectory has a small periapsis distance with the centre of mass of host and perturber.
  Because the orbital elements are calculated at the end of the integration of the particle's trajectory,
  when the distance between host and perturber is usually large again,
  the position of the systems centre of mass may differ considerably from the position of the host.
}
of those particles.

\subsection{Low-mass perturbation}

For the mass-ratio of $m = 0.1$,
the final eccentricities of the unbound particles are shown in the map in a logarithmic colour scale up to $e = 10$
(see Fig.~\ref{fig:m01_5x5_ecc_unbound}).
The particles from the band along the approaching branch of the perturber's orbit finally have high eccentricities,
while the particles from the band along the departing branch have eccentricities slightly above $1$.

Figure~\ref{fig:m01_5x5_periapsis_unbound} shows the final periapsides of the unbound particles.
The particles from the band along the approaching branch of the perturber's orbit 
have finally periapsis distances with the centre of mass comparable to their initial orbital radii.
In contrast, the particles from the departing branch of the perturber's orbit
have relatively small final periapsis distances.
By also taking into consideration the eccentricities of the final orbits of these particles, we can say
that particles with VPPs close to the approaching branch are ejected on hyperbolic orbits with relatively high periapsis distances.
Particles from the departing branch become barely unbound with eccentricities just above $1$ and small periapsides.
For these barely hyperbolic orbits, small periapsides correspond to small semi-major axes and thus to relatively small angular momenta.

\subsection{High-mass perturbation}

For the mass-ratio of $m = 20$, the particles that become unbound finally have remarkably low eccentricities,
most have $e < 2$ (see Fig.~\ref{fig:m20_5x5_ecc_unbound}).
Only particles with VPPs in the upper half, close to the host (cyan region), have final eccentricities $\approx 3$.
Close to the lower border of the image there is a region from where the particles are ejected on orbits with higher eccentricities
(rainbow around e.g. (0,-4.2)).
Compared to the low-mass case, more particles become unbound, but on significantly less hyperbolic orbits.

The gradient in the upper half of the final periapsides of the unbound particles (see Fig.~\ref{fig:m20_5x5_periapsis_unbound})
is caused by the centrifugal clearing (see also Sect.~\ref{sec:high_mass} for explanation).
Only, from the lower half do a few particles become unbound, many of them on barely hyperbolic orbits with small periapsides.

\subsection{Equal-mass perturbations}

\begin{figure*}[t!]
  \centering
  \begin{minipage}[t]{\hsize}
    \centering
        {\bf Mass ratio = 0.1}
  \end{minipage}
  \begin{minipage}[t]{\hsize}
    \centering
    \begin{minipage}[t]{0.42\hsize}
      \vspace{0pt}
      \begin{subfigure}[t]{\textwidth}
        \begin{minipage}[t]{0.05\textwidth}
          \vspace{0pt}
          \caption{}\label{fig:m01_5x5_ecc_unbound}
        \end{minipage}
        \hfill
        \begin{minipage}[t]{0.94\textwidth}
          \vspace{0pt}
          \includegraphics[width=\textwidth]{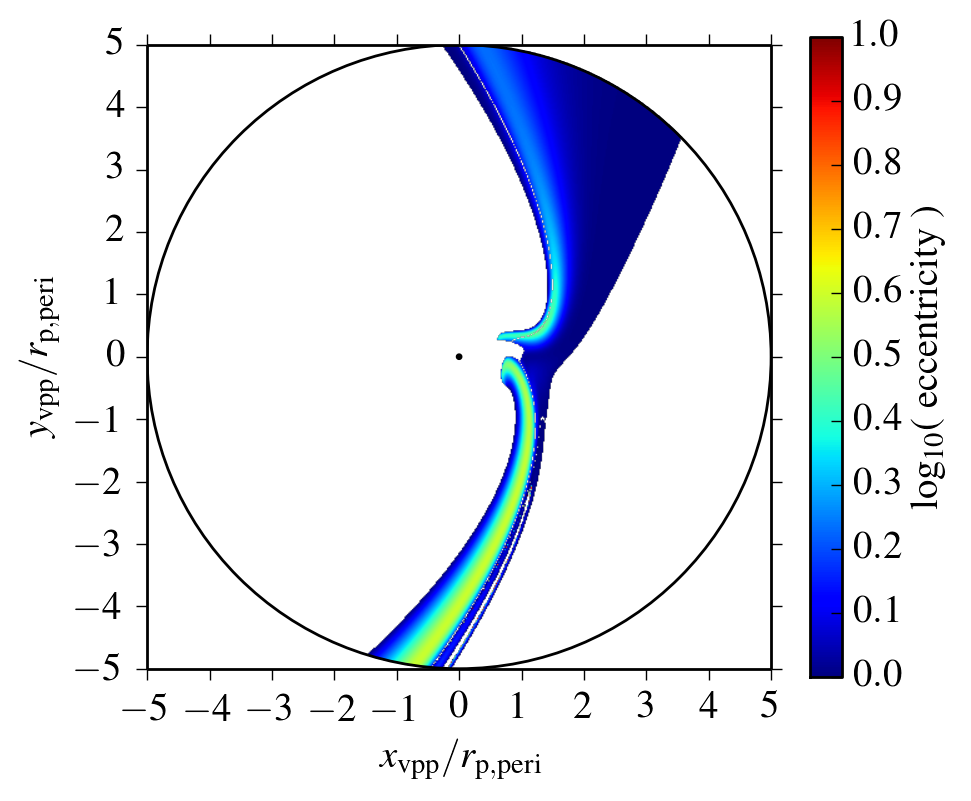}
        \end{minipage}
      \end{subfigure}
    \end{minipage}
    \hspace{1em}
    \begin{minipage}[t]{0.42\hsize}
      \vspace{0pt}
      \begin{subfigure}[t]{\textwidth}
        \begin{minipage}[t]{0.05\textwidth}
          \vspace{0pt}
          \caption{}\label{fig:m01_5x5_periapsis_unbound}
        \end{minipage}
        \hfill
        \begin{minipage}[t]{0.94\textwidth}
          \vspace{0pt}
          \includegraphics[width=\textwidth]{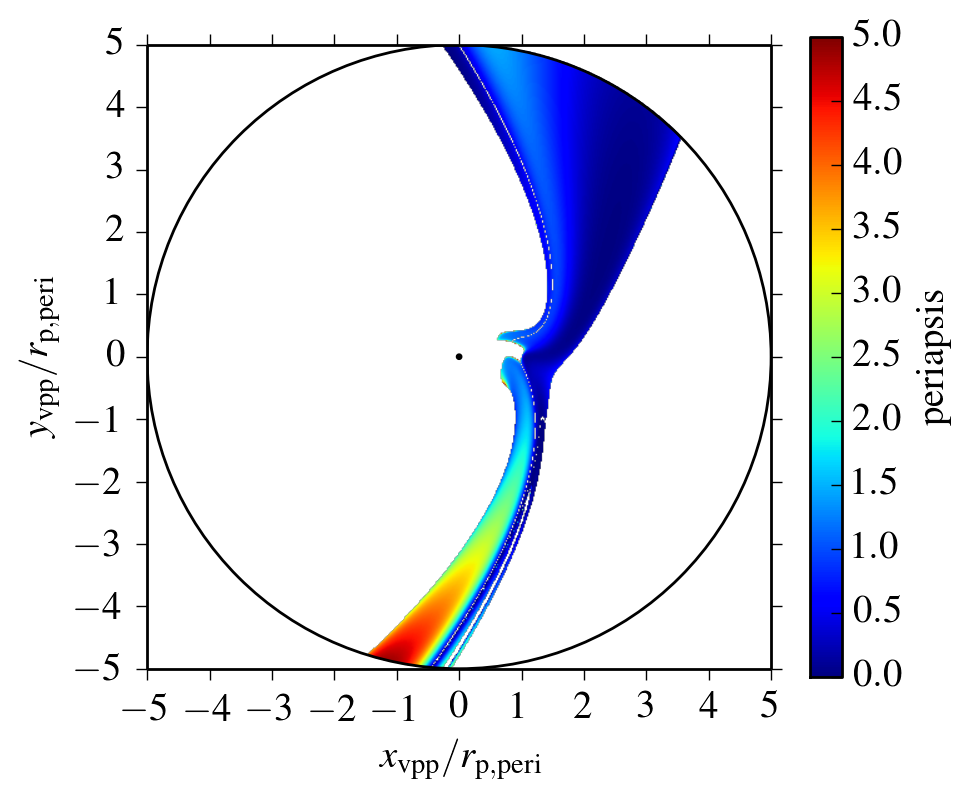}
        \end{minipage}
      \end{subfigure}
    \end{minipage}
  \end{minipage}
  \begin{minipage}[t]{\hsize}
    \centering
        {\bf Mass ratio = 20.0}
  \end{minipage}
  \begin{minipage}[t]{\hsize}
    \centering
    \begin{minipage}[t]{0.42\hsize}
      \vspace{0pt}
      \begin{subfigure}[t]{\textwidth}
        \begin{minipage}[t]{0.05\textwidth}
          \vspace{0pt}
          \caption{}\label{fig:m20_5x5_ecc_unbound}
        \end{minipage}
        \hfill
        \begin{minipage}[t]{0.94\textwidth}
          \vspace{0pt}
          \includegraphics[width=\textwidth]{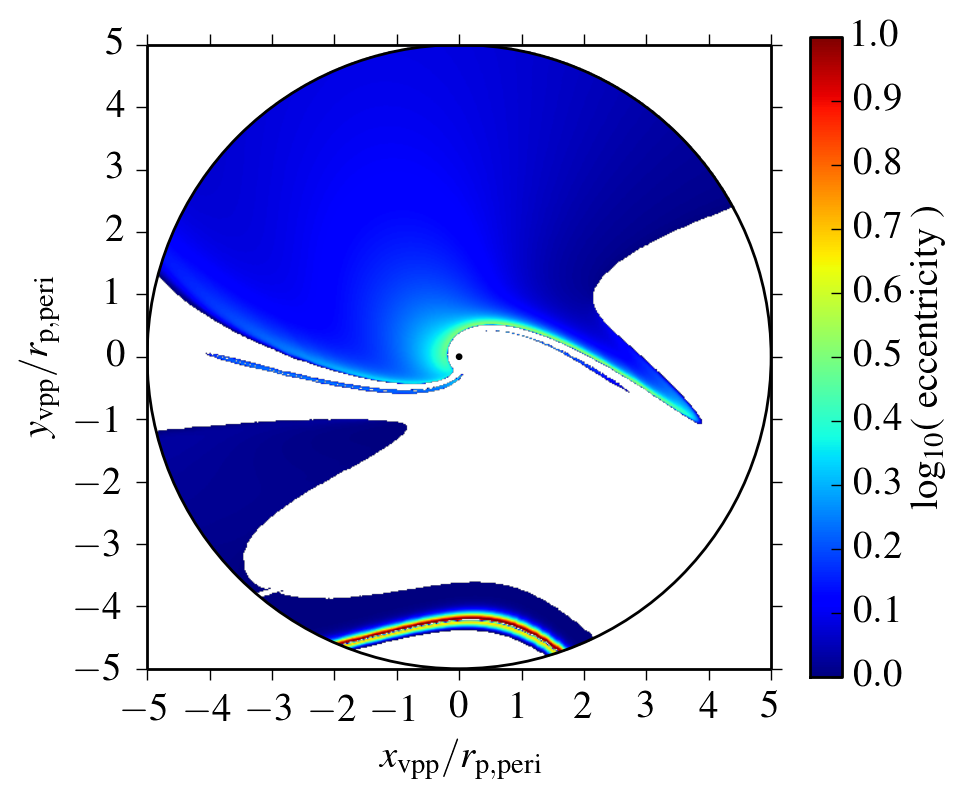}
        \end{minipage}
      \end{subfigure}
    \end{minipage}
    \hspace{1em}
    \begin{minipage}[t]{0.42\hsize}
      \vspace{0pt}
      \begin{subfigure}[t]{\textwidth}
        \begin{minipage}[t]{0.05\textwidth}
          \vspace{0pt}
          \caption{}\label{fig:m20_5x5_periapsis_unbound}
        \end{minipage}
        \hfill
        \begin{minipage}[t]{0.94\textwidth}
          \vspace{0pt}
          \includegraphics[width=\textwidth]{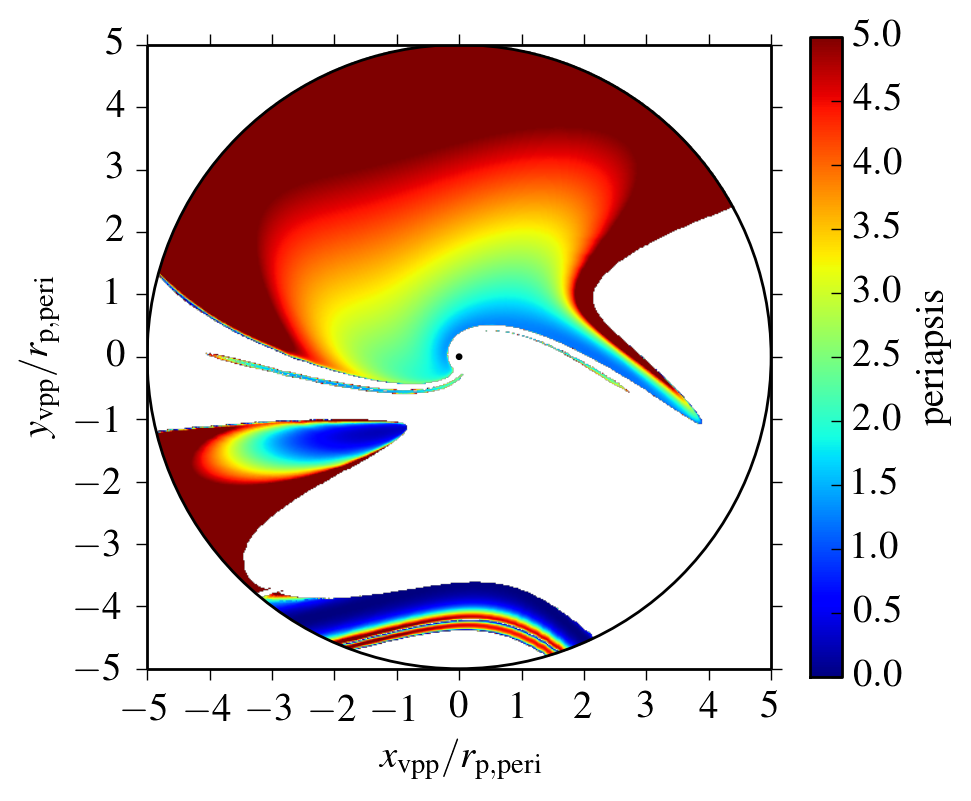}
        \end{minipage}
      \end{subfigure}
    \end{minipage}
  \end{minipage}
  \begin{minipage}[t]{\hsize}
    \centering
        {\bf Mass ratio = 1.0}
  \end{minipage}
  \begin{minipage}[t]{\hsize}
    \centering
    \begin{minipage}[t]{0.42\hsize}
      \vspace{0pt}
      \begin{subfigure}[t]{\textwidth}
        \begin{minipage}[t]{0.05\textwidth}
          \vspace{0pt}
          \caption{}\label{fig:m1_5x5_ecc_unbound}
        \end{minipage}
        \hfill
        \begin{minipage}[t]{0.94\textwidth}
          \vspace{0pt}
          \includegraphics[width=\textwidth]{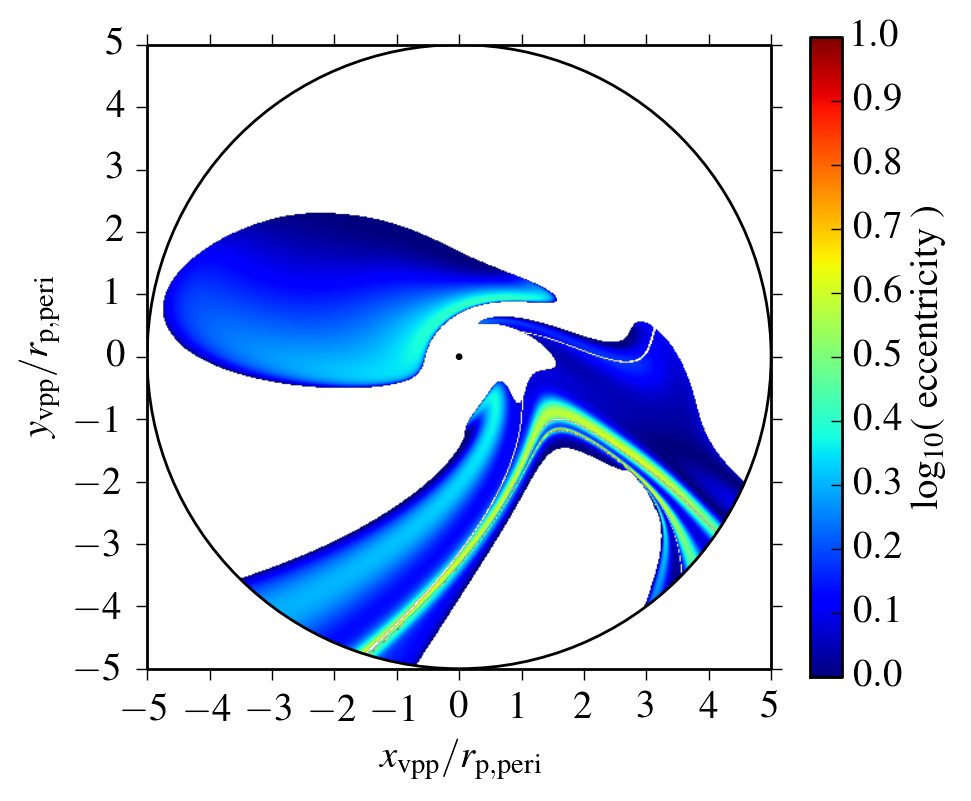}
        \end{minipage}
      \end{subfigure}
    \end{minipage}
    \hspace{1em}
    \begin{minipage}[t]{0.42\hsize}
      \vspace{0pt}
      \begin{subfigure}[t]{\textwidth}
        \begin{minipage}[t]{0.05\textwidth}
          \vspace{0pt}
          \caption{}\label{fig:m1_5x5_periapsis_unbound}
        \end{minipage}
        \hfill
        \begin{minipage}[t]{0.94\textwidth}
          \vspace{0pt}
          \includegraphics[width=\textwidth]{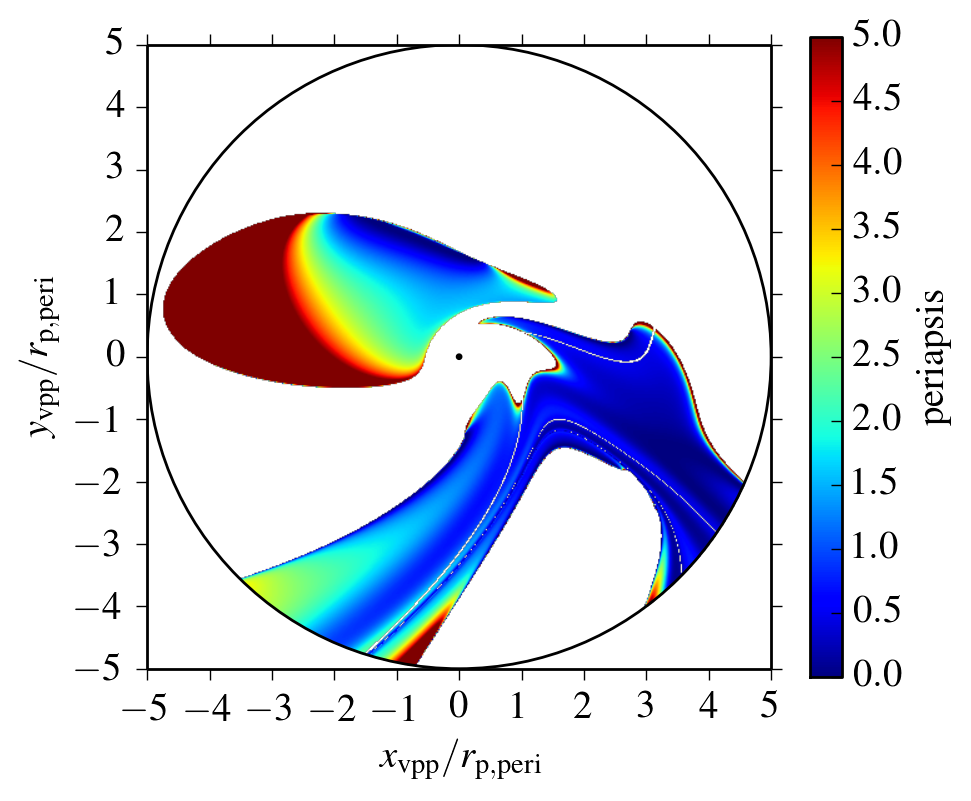}
        \end{minipage}
      \end{subfigure}
    \end{minipage}
  \end{minipage}
  \caption{Maps of the final eccentricities (left column) and periapsides (right column), with the centre of mass
    of the unbound particles for the three mass-ratios.}
  \label{fig:5x5_unbound}
\end{figure*}

The similarities with the low- and the high-mass case, as described in Sect.~\ref{sec:equal_mass}, can also be seen
in the final periapsides with the centre of mass of the unbound particles (see Fig.~\ref{fig:m1_5x5_ecc_unbound}
and Fig.~\ref{fig:m1_5x5_periapsis_unbound}).
In the final eccentricities and periapsides of the particles along the approaching branch of the perturber's interaction orbit,
similarities to the low-mass case can be seen.
The drop-shaped, centrifugally cleared region shows similarities with the high-mass case.
As in the low-mass case, along the approaching branch of the perturber orbit an increase of the final periapsides with 
increasing distance of the VPP to the host can be seen.
A similar increase can be seen in the centrifugally cleared region.
All other unbound particles finally have relatively small periapsides with the centre of mass.
The final eccentricities are again surprisingly low; they are all below $\approx 3$.

\end{appendix}

\end{document}